\documentclass[oldversion]{aa}
\usepackage{graphicx}
\usepackage{caption}
\usepackage{subcaption}
\usepackage{lipsum}
\usepackage{txfonts}
\usepackage{natbib}
\usepackage{mathrsfs}
\usepackage{amsmath}
\usepackage{hyperref}
\usepackage[flushleft]{threeparttable}
\bibpunct{(}{)}{;}{a}{}{,} 

\begin{document}

\title{eROSITA's cool star population explained}

\author{J. H. M. M. Schmitt\inst{1},P.C. Schneider\inst{1}, S. Czesla\inst{1}, S. Freund\inst{1}, J. Robrade\inst{1}}

\institute{Hamburger Sternwarte, Universit\"at Hamburg, Gojenbergsweg 112, 21029 Hamburg, Germany\\
           \email{jschmitt@hs.uni-hamburg.de}}

\date{Received \dots; accepted \dots}

\abstract{

The rotation-activity connection is the standard paradigm for interpreting chromospheric and coronal activity in late-type stars, 
namely, stars with outer convection zones. This paradigm states that activity increases with decreasing rotation period until a saturation
limit is reached. By scaling rotation periods with the convective turnover time via the Rossby number, $\text{Ro}$, saturation is expected 
to occur at a universal value across all spectral types.   In our paper,
we systematically investigate the relationship between rotation and activity as measured though X-ray emission for a large sample of 
late-type stars to test the universal applicability of this paradigm. To this end, we utilized TESS short-cadence space photometry to 
determine the rotation periods for late-type stars identified in the eROSITA all-sky survey. This combined dataset provides rotation and 
X-ray measurements for 14004 stars, representing a sample size increase of more than an order of magnitude compared to 
previous studies. Our results show that F-type stars do not reach the "classical" saturation limit of L$_X$/L$_{\text{bol}} \approx 10^{-3}$, instead exhibiting significantly lower $L_X/L_{\text{bol}}$ values. We find that the convective turnover times derived from this sample closely agree with theoretical 
computations, supporting the idea that Rossby number-activity relations hold for all late-type main sequence stars. The lower level of activity in 
earlier spectral types (e.g., F-type and late A-type stars) is a physical consequence of their short convective turnover times, which 
prevent them from rotating rapidly enough to ever reach the saturation regime. We demonstrate that a simple model incorporating our 
derived turnover times versus color can successfully reproduce the observed characteristics of the eROSITA X-ray activity distribution, as 
measured by the L$_X$/L$_{\text{bol}}$ ratio and {\it Gaia} BP-RP color.}

\mail{J.H.M.M. Schmitt, jschmitt@hs.uni-hamburg.de}
\titlerunning{Coronal activity of late-type stars}
\keywords{Stars: activity; Stars: coronae; Stars: late-type}

\maketitle
\nolinenumbers

\section{Introduction}
\label{sec_intro}

The most fundamental stellar parameter controlling the activity level of a given late-type star is its
rotation rate.   This leads to the rotation-activity paradigm, which roughly states that rapid rotators
are the most ''active'' stars with the largest surface coverage of star spots and the highest levels of
chromospheric and coronal emission.   In this paper, we are concerned with
photospheric modulations caused by star spots, which are used to
provide estimates of stellar rotation periods as well as
with coronal emissions observed at X-ray wavelengths.
Following the pioneering study of \cite{pallavicini1981}, numerous studies of the relation 
between X-ray emission and rotation periods of stars have been carried out; a recent review of this topic is 
provided by \cite{santos2024}.
These studies have contributed towards making
the rotation-activity paradigm the commonly accepted standard framework to discuss and interpret 
all magnetic ''activity'' phenomena observed in late-type stars.  It is important to keep in mind that
''activity'' phenomena in late-type stars
comprise a plethora of different observational findings observed at different
wavelength ranges and typically referring to different parts of the outer atmospheres of
the underlying stars; for a recent review, we refer to the textbook by \cite{basri2021}.

Star spots are among the best tools presently at our disposal to determine the rotation period of a given star.
To the extent that the star spots are sufficiently inhomogeneously distributed on the stellar surface and 
''live'' sufficiently long, the periodic signatures in the resulting stellar light curves are directly
related to the sought for stellar rotation period.  Needless to say, nature can be unkind at times
and we refer to \cite{2020basri} for an in-depth study of the information content of 
photometric light curves.
The availability of space-based high-precision photometry has overcome the severe limitations of
ground-based photometry in terms of temporal coverage and photometric precision;
for example, using data from the {\it Kepler} satellite, \cite{mcquillan2014} were able to measure
periods for more than 34000 stars.  Yet, as was made apparent by the detailed mission description by \cite{borucki2010},
the sky area covered by the {\it Kepler} satellite was 
limited and the target stars are usually too faint to be detectable as X-ray sources even in deep
X-ray pointings.

Stellar coronae can be best diagnosed in the X-ray range; since stellar photospheres are
X-ray dark, an X-ray flux measurement provides a direct measure of the coronal emission strength
of a given star.   Thanks to eROSITA's all-sky survey \citep{merloni2024} the number of
known X-ray sources and in particular X-ray emitting stars has increased tremendously in recent years.
Already the first half year of eROSITA's all-sky survey \citep{merloni2024} has 
provided $\approx$  1.3 million X-ray sources detected in the 0.2~--~2.3~keV energy band  \citep{merloni2024},
and more than 100000 of these X-ray sources can be identified  with so-called normal stars, the majority being
cool dwarf stars \citep{freund2024}.  Since X-ray surveys tend to focus on the more X-ray luminous active stars, 
according to the rotation-activity paradigm, we would expect the underlying population of X-ray-detected stars to
consist of predominantly rapid rotators.

Thus, it is clear that to understand the population of cool stars associated with eROSITA X-ray sources, we
 need rotation periods for many thousands of stars distributed over the sky, with the expectation that
many of these periods are rather short and the X-ray emission is at or close to the saturation limit.
As shown by  \citep{freund2024}, the majority of the expected stellar counterpart population
is (by comparison) optically bright (i.e., brighter than 13-15 in the {\it Gaia} G-band, depending on spectral type).
 A valuable source for rotation periods of such stars is provided by
the  Transiting Exoplanet Survey Satellite (TESS) mission, which provides short-cadence photometry over the whole sky
for brighter stars \cite{ricker2015} and it is therefore close to ideal to be used in conjunction with the eROSITA X-ray all-sky survey.
By combining eROSITA and TESS, we ought to be able to extract such data for many thousands of
stars. At the same time, accurate {\it Gaia} photometry is available for almost all our sample stars, allowing for a
precise placement of the stars in a color-magnitude diagram and obtaining an estimate of stellar mass for
main sequence (MS) objects.

The other important quantities that are useful for  characterizing the activity of late-type stellar X-ray sources is the convective turnover time.
\cite{noyes1984} were the first to introduce the Rossby number (i.e., the ratio between rotation period and
convective turnover time) into the interpretation of stellar activity data. They derived
a unique period-activity relationship for their sample of 
chromospherically active late-type MS stars 
and demonstrated  a rather tight relationship between rotation and activity:
stars with small Rossby numbers (i.e., with short rotation periods) exhibit high levels of (chromospheric) activity and vice versa; as pointed out by \cite{noyes1984},
this finding suggests an interpretation in terms of dynamo related activity
since simple scaling relations yield a dependence of $N_D \sim R_{o}^{-2}$, with $N_D$ being the 
dynamo number, characterizing the ``dynamo strength."
The approach of \cite{noyes1984} was carried into the X-ray range by
\cite{2003pizzo} and \cite{2011wright}. They considered
how, for example, the X-ray luminosity, $L_X$, or the fractional X-ray luminosity,  L$_X$/L$_{bol}$, depend on
rotation, using stellar samples with known X-ray properties and rotation periods.
In the X-ray range, \cite{2011wright} found X-ray ''saturation'' at a level of  $L_X/L_{bol}$ $\approx$ $10^{-3}$ for
Rossby numbers below 0.13, while for larger Rossby numbers (i.e., slower rotators), the  $L_X/L_{bol}$ ratio
decreases rapidly with an increasing Rossby number.

The convective turnover time determinations by \cite{noyes1984}, \cite{2003pizzo}, and \cite{2011wright} were all
entirely empirical; whereas, in contrast,  \cite{landin2010} and more recently \cite{landin2023} published theoretical estimates of 
convective turnover times for cool stars.
Specifically, \cite{landin2023} published convective turnover times for stars covering the mass range 0.1 - 1.5 M$_{\odot}$ for various 
evolutionary stages. In particular, these authors produced both global convective turnover times defined as the means over the whole stellar convection zone,
along with local convective turnover times computed at half a mixing length above the base of the convection zone; we
refer to Sects. 3.1 and 3.2 of \cite{landin2023} for details.   As \cite{landin2023} have shown, there is reasonably good agreement 
between the empirical and theoretical values for convective turnover time; hence, the introduction of the Rossby number appears to be a
sound concept. 

The purpose of this paper is to analyze the available TESS short-cadence data for those
sources  identified with coronal emitters from the eROSITA all-sky survey to derive their periods, whenever possible.
To focus our paper on the physical results, we moved all sections primarily concerned with technicalities into the
appendices. We provide on a brief overview of the adopted procedures in Sect.~\ref{sec_to_meas},
while the bulk of our paper deals with the derived results and their interpretation. The specific plan of our paper is as follows. 
In Sect.~\ref{sec_to_meas} we first provide a brief overview of our
TESS period determinations, while the more technical details of our work are presented in Appendices \ref{app:sec_abavdor} and \ref{app:sec_per_det}.   The resulting periods are also presented in
Sect.~\ref{sec_to_meas}, as well as a comparison with other period determination efforts, and we also give the statistics and 
reliability of our period determinations in comparison.
The astrophysical results of our work are presented in Sect.~\ref{sec_result_disc}, where we describe the construction of
our reference stellar sample.  We also present color-magnitude diagrams of the stellar sources and study the dependence
of activity on rotation periods.   We introduce the Rossby numbers and derive the convective turnover times from our stellar sample.
We show that rapidly rotating stars of spectral type F6 or earlier cannot reach the saturation regime. Finally, we present a
simple toy model that describes the X-ray activity of eROSITA's cool star population as a function of color.

\section{TESS rotation periods of eROSITA detected coronal sources}
\label{sec_to_meas}

In this paper, we consider X-ray sources extracted from the 1.3 million X-ray sources detected in the 0.2~--~2.3~keV energy band during the
first half year of eROSITA's all-sky survey \citep{merloni2024}.   Specifically, we used our HamStar algorithm \citep{freund2024} to identify 
stellar (coronal) sources. 
In brief, HamStar uses information from {\it Gaia} DR3 \citep{gaia2023} and Tycho \citep{hog2000} to calculate the 
probability, $p_{stellar}$, that a given eROSITA X-ray source is actually  a coronal X-ray emitter and we refer to   \cite{freund2024}
for a detailed account of the adopted identification procedures.  For our study, we consider all eRASS1 sources 
with $p_{stellar}>0.5$ and a  {\it Gaia} DR3 ID as an identification.   The so-called TESS input catalog (TIC; \citealt{tic2022})
provides  {\it Gaia} DR2 IDs for most sources and we mapped the {\it Gaia} DR3 IDs from HamStar to 
 {\it Gaia} DR2 IDs using the \texttt{dr2neighbourhood} table available as part of  {\it Gaia} DR3\footnote{Some of our sources have more 
than one possible DR2 counterpart.  In those cases, we selected the one closest in magnitude to the DR3 value.}. 
In this way, the high probability stellar
eROSITA X-ray sources were associated with a TIC~ID using this  {\it Gaia} DR2 ID and 
a total of about 69,000 sources resulted from this procedure. 

The primary scientific goal of the TESS mission is the detection of exoplanets around the brighter stars by
performing differential photometry  with a cadence of
two minutes simultaneously, utilizing its
four red-band wide-field cameras that cover
a 24$^{\circ}$ $\times$  96$^{\circ}$ strip on the sky for a total 27 days (denoted as ''sector'' in the TESS
terminology). A detailed description of the TESS mission is given by \cite{ricker2015}. 
We checked which sources had been observed by TESS in short-cadence mode in sectors 1 - 58, downloaded the
respective data from the TESS-MAST archive, and analyzed the data following the procedures described in Sect.~\ref{app:sec_per_det}.
In addition to the {\it Gaia} data, HamStar also incorporates information from Tycho and some eROSITA sources, mainly the optically brighter ones, 
currently have only a Tycho but no {\it Gaia} counterpart. For those sources, 
we found that matching based on Tycho ID left many 
sources without a reasonable TIC counterpart and we opted for a distance based matching for those sources, which 
added 375 sources to our input catalog.  Thus, in total, we obtained 69179 eRASS1 sources in the German half of the eROSITA sky 
brighter than G $=$ 13  and contained in the TESS input catalog.   

The TESS data of a given source do not come as a contiguous data stream even in a single TESS sector;
rather the data stream is interrupted, usually after about 13~days, when the
data taken during one TESS orbit is downloaded to the ground.   
Furthermore, the TESS data for a given object
might show pronounced jumps between the first data segment in a given sector and
the second segment, which is clearly instrumental.   These gaps and instrumental effects 
obviously are a challenge for the data analysis.   In addition, the coverage of sources with respect to
TESS sectors is quite inhomogeneous.    Sources near the ecliptic poles receive a lot
of exposure, those near the ecliptic equator much less, while periods derived in different sectors
may agree or disagree and, thus, a special treatment is required.

Measuring a period in a noisy data set can be quite cumbersome.   To reduce
possibly incorrect determinations of periods we used three different
procedures to assess whether significant periodicity is contained in a
given data set.   Specifically, we use the generalized Lomb-Scargle  (GLS)
analysis, as developed by \cite{ferraz1981} and
\cite{zech2009}, the autocorrelation function of the light curve data following the
approach by \cite{mcquillan2014}, and the 
phase dispersion minimization procedure developed by \cite{stelling1978}.
These three methods measure somewhat different properties of the analyzed
data stream:  the GLS analysis attempts to identify the most significant Fourier
components in the data, the autocorrelation approach investigates the
correlation properties of the data, and the phase dispersion minimization seeks
those periods which result in the ''best'' light curves in the sense
that the folded data show the least amount of dispersion around some ''mean'' light curve.

The rotational signals in these eROSITA-selected TESS sources are typically not produced
by eclipses (although we do find a fair number of eclipsing binaries
in our sample); rather, they are produced by star spots inhomogeneously
distributed on the stellar surface.    However, star spots come and go, their
surface distribution changes (maybe all the time); furthermore, the stars may
rotate differentially, thus  all these effects may change the rotational
signatures in the data.  On the other hand, the light changes in an
eclipsing binary are very regular, except that the surfaces of such stars are often also
covered by time variable spots.

\begin{figure}  [h]
\centering
 \includegraphics[width=8cm]{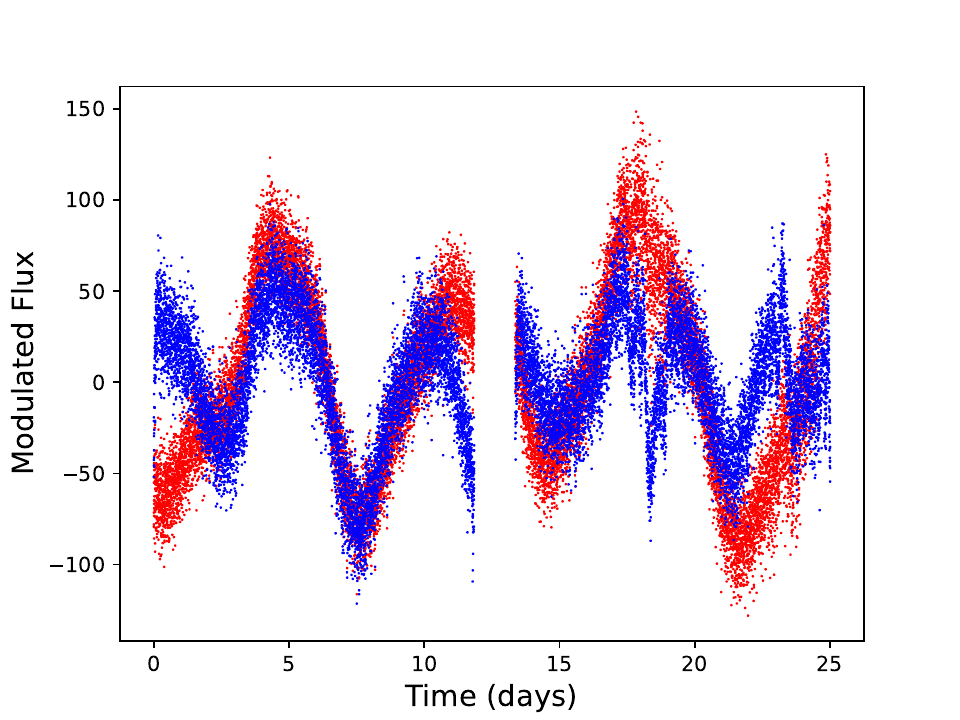}
   \caption{Comparison of TESS SAP light curve (red data points) and TESS PDCSAP (blue data points) for
   the source  TIC~270771392 observed in sector 34;  see text for details.}
\label{comp_sap_pdcsap}
\end{figure}

\subsection{Period determination:  Overview}
\label{sec_period_det}

In this section, we provide an overview of our period determination scheme, with more detail found in Appendix~\ref{app:sec_per_det}.   The TESS data downloaded from the TESS-MAST archive 
contain two types of light curve files: one with fluxes based on
Simple Aperture Photometry (SAP) and one with fluxes based on pre-search data conditioning SAP (PDCSAP).
The PDCSAP algorithm attempts to identify and correct for systematic instrumental trends in the
TESS light curves and is geared towards transit detection;
for a detailed discussion of the applied procedures, we refer to \cite{jenkins2016}.
 However, we  found TESS light curves where an obvious overcorrection
was performed by the PDCSAP procedures.  An example is shown in Fig.~\ref{comp_sap_pdcsap}, which shows the SAP and
PCDSAP light curves in comparison for the case of TIC~270771392 observed in sector 34. We note that only the modulated part of
the light curve  is shown and a constant has been subtracted.   As can be seen from  Fig.~\ref{comp_sap_pdcsap}, the two
light curves agree well in the time range of 3-10~days and 14-16~days, yet the start and end of the first observing interval
and the features near  days 16-18 and day 23 until the end of the observations are quite different and actually lead to incorrect
period estimates.   In the specific case of TIC~270771392, another light curve was  obtained in sector 8, which
suggests that the SAP values are much closer to the truth than the PDCSAP values;   other examples
of apparent overcorrections by the PDCSAP procedure are given by \cite{hedges2021}.

We therefore decided to work with SAP fluxes and applied only some moderate light curve corrections:
sectors with more than one data gap of one day or more were ignored altogether to avoid additional light curve aliasing, while the 
data in the two sectors halves were adjusted to the same mean to avoid long-term trends. Finally, we computed the dispersion 
$\sigma$ of the light curve and rejected all data with deviations of more than three $\sigma$ to reduce the influences of
systematic trends that are often seen at the beginning or end of TESS sectors.   It is clear that we do not claim to have removed
all instrumental effects from the light curves and from our results (presented in Sect.~\ref{stell_samp}) and some instrumental effects
will affect these periods.

To deal with all these effects, we devised a special treatment of the derivation of TESS rotation periods.
Our period determination scheme is primarily based on the Lomb-Scargle 
periodogram power (GLS; as defined by \citealt{zech2009}); however, we also considered
phase dispersion minimization and/or autocorrelation to finally arrive at a period assignment for every
X-ray source; a detailed account of the adopted procedures is given in Appendices~\ref{app:sec_longper} and~\ref{app:sec_shortper}.
To indicate our confidence in these period assignments we introduced an empirical
quality grading scheme ranging from grades 1-7.   A period grade of 4 requires a consistent period
determination of the data in a given sector (to within 10\%) with three different period searching methods.
If successful and consistent period measurements are available in more than a single sector,
the assigned period quality grade is increased by one, if at least three consistent period measurements 
in different sectors are available, by 2 and so on.  The maximally possible grade is
7, implying that there are at least nine consistent high quality period measurements available for the given 
source.   A more detailed description of this procedure is given in Sect.~\ref{app:sec_fin_per},
where we also provide a table with the derived grade frequency distribution.  
For 14004 sources, we managed to obtain valid period measurements with grades 3 and larger, whereby
11731 of the derived periods have quality assignments greater than 3.  That does not 
imply that the periods for the remaining 2273 sources are wrong; however, the rate of incorrect period determinations
among these sources is expected to be  higher.

An exemplary listing of our period results is given in Table~\ref{tab1}, the full table is available in the online material. In Table~\ref{tab1}, we provide the respective TIC-ID numbers, the derived periods and their errors (whenever available)
The flag MS denotes MS stars, the flag PB indicates a "possibly bad"\ period, namely, where we suspect it is not a rotation period of the
target object. For more details, we refer to Sect.~\ref{stell_samp}.

\begin{center}
\begin{table}
\caption{\label{tab1} Successful period determinations (extract).}
\begin{tabular}{rrrrrr}
\hline
  TIC-ID &  Period &  Period error &  Grade & MS & PB \\
             & (day)     & (day)             &             &flag & flag\\
\hline
102723 &  4.05 & 0 & 4 & 1 & 0 \\
113645 & 12.79 & 0 & 4 & 1 & 0 \\
120016 &  0.16 & 0 & 4 & 1 & 0 \\
122797 &  7.39 & 0 & 4 & 0 & 0 \\
138893 & 15.13 & 0 & 4 & 1 & 0 \\
141778 &  3.44 & 0 & 4 & 1 & 0 \\
155455 &  3.17 & 0 & 3 & 0 & 0 \\
167121 & 10.35 & 0 & 4 & 1 & 0 \\
592590 &  8.35 & 0 & 4 & 1 & 0 \\
 593228 & 14.76 & 0.23 & 5 & 1 & 0 \\
 610232 &  4.98 & 0 & 4 & 1 & 0 \\
 614576 & 14.91 & 0 & 4 & 1 & 0 \\
 615133 &  0.86 & 0 & 4 & 1 & 0 \\
 665581 &  0.24 & 0 & 4 & 1 & 0 \\
 671564 &  0.19 & 0 & 4 & 1 & 0 \\
 672624 &  0.56 & 0 & 4 & 1 & 0 \\
 675445 & 10.72 & 0 & 4 & 0 & 0 \\
 710947 &  9.05 & 0 & 4 & 1 & 0 \\
 737327 &  1.62 & 0 & 5 & 1 & 0 \\
 768859 &  8.85 & 0.44 & 5 & 1 & 0 \\
 775434 &  5.64 & 0.02 & 5 & 1 & 0 \\
 862763 &  1.01 & 0 & 3 & 1 & 0 \\
 864120 &  8.10 & 0 & 4 & 1 & 0 \\
 864246 &  7.67 & 0 & 4 & 1 & 0 \\
 864738 &  1.46 & 0 & 3 & 1 & 0 \\
 893123 &  8.77 & 0 & 4 & 1 & 0 \\
 990827 & 10.30 & 0.21 & 5 & 1 & 0 \\
 990903 &  8.03 & 0.29 & 5 & 0 & 0 \\
\hline
\end{tabular}
\tablefoot{The full table is available at the CDS. Col. 1 gives
TIC ID number. Col. 2 the derived period. Col. 3 indicates its error. Col. 4 provides the quality flag. Cols. 5 and 6 give the MS
flag (i.e., objects in the red polygonal
area in Fig. 6) and PB flag (i.e., bad period objects inside
green dashed area in Fig. 7). We note that a period error is computed only
for sources with successful period determinations in at least two sectors.}
\end{table}
\end{center}

\begin{figure}  [t]
\centering
 \includegraphics[width=8cm]{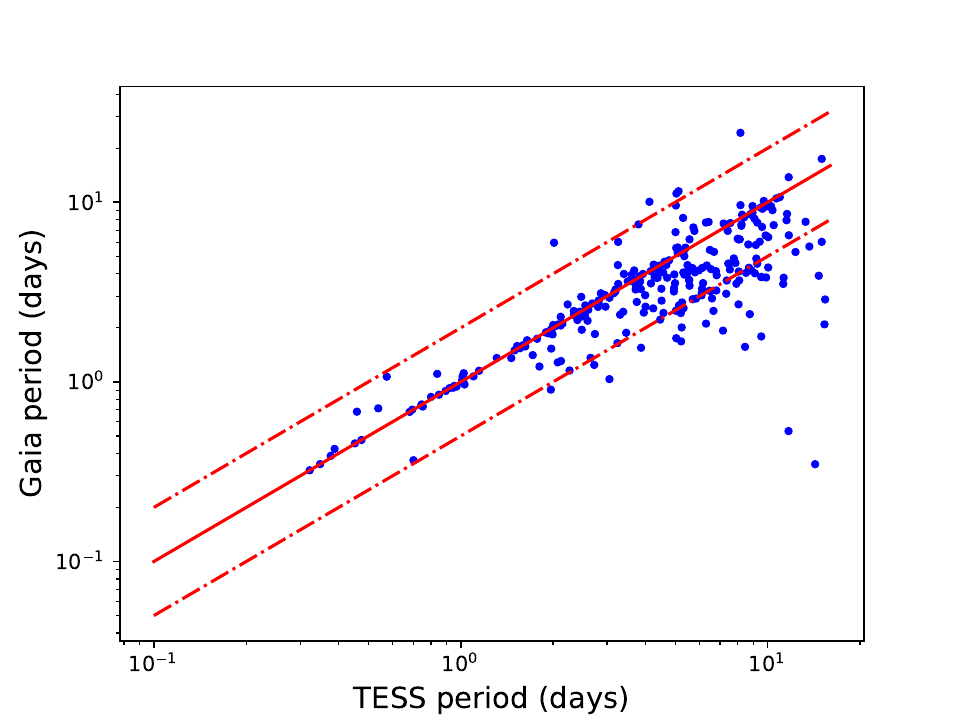}
   \caption{Comparison of TESS periods as derived in this paper and {\it Gaia} periods 
as derived by \cite{distefano23} for 275 objects with eROSITA detections. Solid line
indicates consistent period determinations, dash/dotted lines indicate a period ratio
of 2 and 1/2;  see text for details.}
\label{comp_tess_gaia}
\end{figure}

\subsection{Comparison with other period determinations}
\label{sec_period_comp}

We first compare our new period determinations with other 
period determination efforts and check to what extent the different applied methodologies
agree or disagree.   For this sanity
check, we used the rotation modulation studies of  {\it Gaia} sources published by
\cite{distefano23}; the study of the relation of X-ray activity and rotation for 
cool stars by \cite{2011wright}, who used a variety of ground-based rotation measurements for 824 known
X-ray sources associated with cool stars; the results of
\cite{prsa2022}, who presented a catalog of 4584 eclipsing binaries detected in TESS short-cadence 
observations; and the
study by \cite{mowlavi2023}, who presented a catalog
of more than 500000 eclipsing binary candidates based on {\it Gaia}  data.
Finally, we compared our periods to those produced using the SpinSpotter code developed
by  \cite{holcomb2022}.

\subsubsection{Comparison with {\it Gaia} periods}

Using the brightness measurements of the various focal plane transits observed 
during the {\it Gaia} mission, \citet{distefano23}  analyzed the observed variations and
present determinations of rotation modulation for about 474\,000 {\it Gaia} DR3 late-type stars. 
Due to the complex {\it  Gaia} scanning law, the sampling properties and the completeness of
this sample vary considerably over the sky.  \citet{distefano23} argued that
the detection of rotation periods below about five days is favored and estimated
that about 70 to 80~\% of the detected periods are correct.   Upon inspecting
the brightness distribution of those sources with reported periods, we find that
most of these stars are below the TESS brightness limit; however, 
a small fraction of  275 objects
has an eROSITA detected counterpart and a TESS light curve with a
successful period measurement with grade 4 or better.

In Fig.~\ref{comp_tess_gaia}, we compare ''our'' TESS period measurements obtained from
TESS data (as described above) to the periods determined from  {\it Gaia} data as reported by  \citet{distefano23}.
In Fig.~\ref{comp_tess_gaia}, the red solid line indicates period equality, while the dashed lines indicated
aliases by factors of 1/2 and 2.  We specifically find 129 sources with a period agreement
of better than 15\% and another 38 sources where the {\it Gaia} period is twice the TESS period,
which could be explained by an alias of the TESS period measurement.

We checked the largest two outliers (with the TESS period above 10~d,
and {\it Gaia} period below 1~d).   In the case of TIC~139285669, we conclude that the TESS period of
14.3~d is instrumental; the {\it Gaia} period of 0.35~d is present in the TESS data as well; however it is above our cutoff of 0.1 in GLS power only
in one of three sectors. Similar considerations apply to
TIC~459944057, which was observed only in one sector; the {\it Gaia} period of 0.53~d is clearly
present in the TESS data (but below our chosen cutoff) and the long period is of instrumental origin.

\begin{figure}  [htb]
\centering
 \includegraphics[width=8cm]{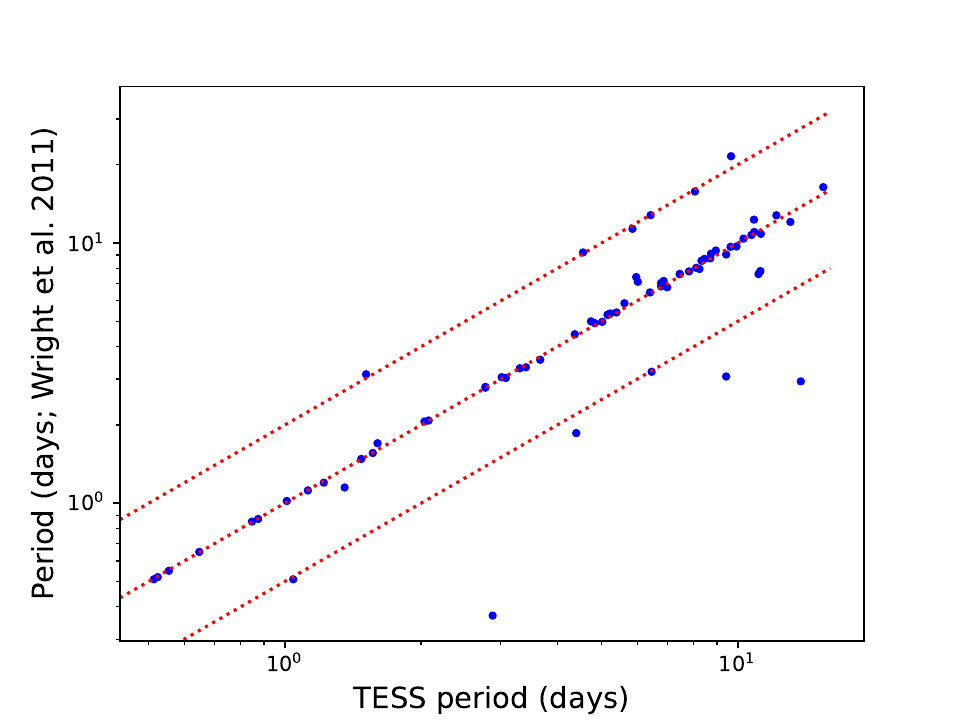}
   \caption{Comparison between rotation periods of cool stars determined by \cite{2011wright} and those in this paper. The unity line as well 
   as the lines for half and twice the periods are also indicated. See text for details.}
   \label{comp_tess_wright}
  \end{figure}

\subsubsection{Comparison with periods derived by \cite{2011wright}}

\cite{2011wright} present a study of non-simultaneous X-ray and rotation measurements of 824 cool stars with an emphasis on the
rotation-activity relation and its dependence on spectral type.   The rotation measurements presented by  \cite{2011wright}  cover periods
between about 0.2 d up to 40 d (i.e., a period range not covered by our analysis).   Obviously, the long-period systems are expected to
exhibit lower X-ray activity and are therefore not expected to show up in shallower all-sky surveys.  
Again considering only matches with grade 4 or better, we find 74 matches and 
show the comparison between the periods  measurements in Fig.~\ref{comp_tess_wright}.   For 57 sources, the measured period ratios
are within 15\%, while another eight cases can be attributed to aliasing; thus, the agreement appears to be very satisfactory.
There is one significant outlier due to the star TIC~455007634; an even casual inspection of the TESS data shows that the period of 0.37~d listed
by \cite{2011wright} cannot possibly be correct.

\begin{figure}  [htb]
\centering
 \includegraphics[width=8cm]{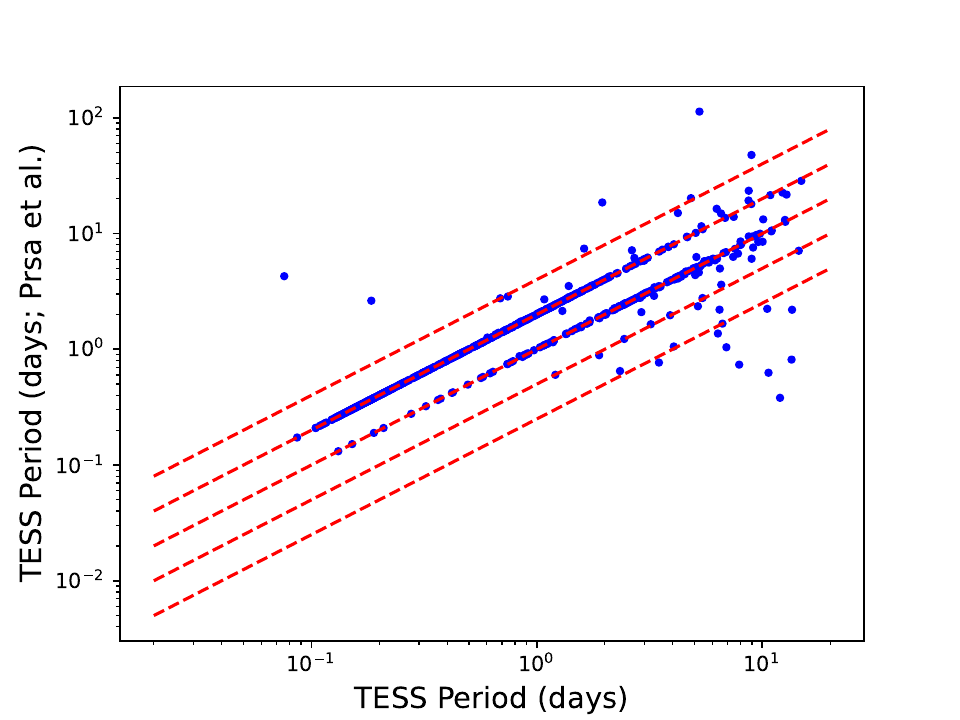}
   \caption{Comparison of periods derived in this paper with those derived by \cite{prsa2022} for a sample of eclipsing binaries.   The unity line is shown as well lines
   at 1/4, 1/2, 2, and 4 times the period. For details, see text. }
   \label{comp_tess_prsa}
  \end{figure}

\subsubsection{Comparison with periods derived by \cite{prsa2022}}

\cite{prsa2022} present a catalog 4584 eclipsing binaries identified in TESS sectors 1-26 in the basis of short-cadence observations. For a detailed description of
the detection and validation procedures, we refer to the discussions in \cite{prsa2022}.  

Among the eclipsing binaries 
cataloged by \cite{prsa2022}, we find an eROSITA X-ray source with a TESS period for more than 706 entries and show the comparison in 
Fig.~\ref{comp_tess_prsa}.   For 163 entries, the periods agree within 15\%;
for 494 entries., \cite{prsa2022}  found twice our period (to within 15\%); and for eight sources, it is the other way around.
Thus, if we include the aliased sources, the agreement is very good, however, it is
also very clear that for eclipsing binaries the bias, for example  demonstrated in 
Fig.~\ref{app:avdor_glsauto}, is very important and Fig.~\ref{comp_tess_prsa} suggests that our procedure, which is based on
GLS periodograms and not optimized for eclipsing binaries, has a strong tendency to pick up half the binary period which is
not surprising.

\begin{figure}  [htb]
\centering
 \includegraphics[width=8cm]{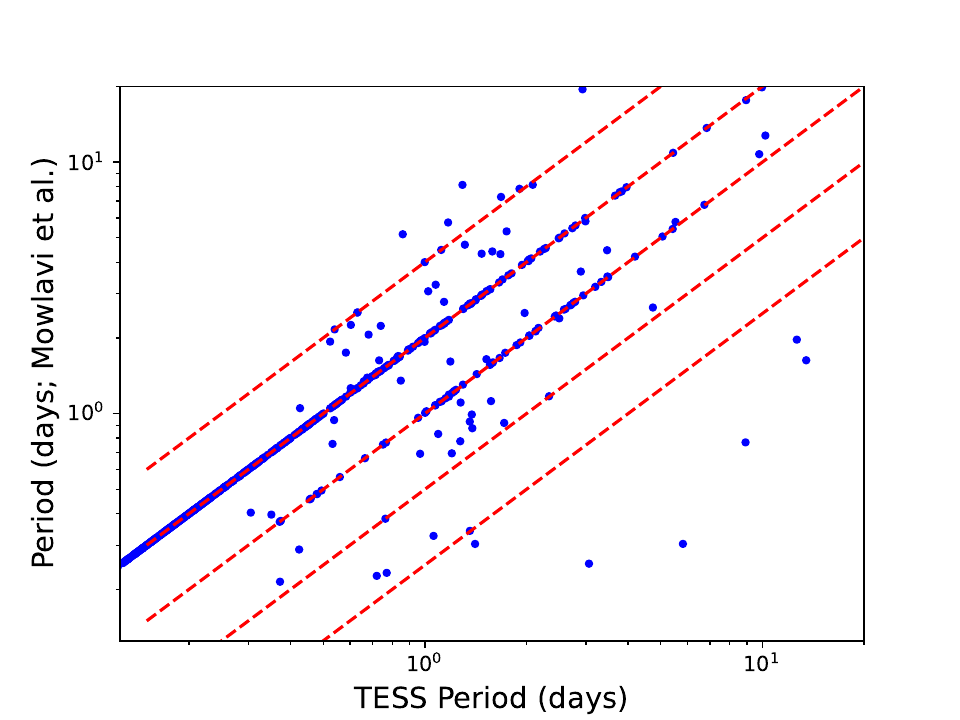}
   \caption{Comparison of periods derived in this paper with those derived by \cite{mowlavi2023} for a sample of eclipsing binaries.   The unity line is shown as well lines
   at 1/4, 1/2, 2 and 4 times the period; for details see text. }
   \label{comp_tess_mov}
  \end{figure}

\subsubsection{Comparison with periods derived by \cite{mowlavi2023}}

\cite{mowlavi2023} present a catalog of more than 500000 eclipsing binary candidates based on {\it Gaia} photometry.   
Again, many of these
sources are quite faint and therefore out of the reach of both TESS and eROSITA.   Nevertheless we find a match for
622 of these sources and show a period comparison in Fig.~\ref{comp_tess_mov}. We note that in this case 
the origin of the data for the
period determinations are different, namely, TESS for ''our'' periods and {\it Gaia} for those presented by \cite{mowlavi2023}.   

A detailed examination shows that the period ratios agree to within 15\% in 63 cases, while in 498 cases, we find half the
period (to within  15\%) of those derived by \cite{mowlavi2023} and in 3 cases, it is the opposite case.  Thus, including aliased
periods the agreement again is satisfactory, but again, as is the case for the sample presented by  \cite{prsa2022}, the
tendency for GLS to pick up half the binary period is very  obvious.

We specifically checked a data point which yielded a large TESS period (3.06~d) and small {\it Gaia} period (0.25~d), which refers to the
star TIC~142742998 (= BD-03 2599), a known, but little studied spectroscopic binary of spectral type G8e; the TESS light curve shows 
two eclipses and a photometric modulation with a period of slightly over 3 days.  Therefore, we are very confident that ''our'' TESS period is correct.

\subsubsection{Comparison with periods derived by \cite{holcomb2022}}

Using TESS data, \cite{holcomb2022} derived periods for 13504 stars out of a sample of 136,868 objects with their SpinSpotter code, a 
period search code based on autocorrelation analysis.   Unfortunately, the results of  \cite{holcomb2022} are not publicly available, yet the
SpinSpotter code is.   To compare to our results,
we used the SpinSpotter code on the SAP fluxes---noting that
\cite{holcomb2022}  used PDCSAP fluxes---for 946 stars  with TESS data available for two or more sectors, periods of less than 20 days, and
a relative dispersion of less than 0.3.  In this sample, 85.5\% of the objects have the same period, whereas in 5.8\% of the cases, the
SpinSpotter periods are twice as large and in 0.7\% of the cases, the opposite is the case.   Thus, if we include the aliased
periods, we can find an agreement in more than 90\% of the cases considered.

\subsubsection{Summary of period comparisons}

Overall, we consider the agreement between previously derived periods and our new TESS-based periods more than satisfactory.   It is clear that in the
case of eclipsing binaries our GLS-based period determination has a considerable probability to pick up only half the period similar to the case of
an active star with two centers of activity on opposing hemispheres.    There is (as usual) the problem of aliasing, which is difficult to ascertain in a 
statistical way.   It is clear that some of our TESS periods are of instrumental nature in particular those near about 13~days;  however, the comparison
with other period determination efforts that do not suffer from this 13-day bias inherent in the TESS data shows that in quite a few cases, our long periods
(i.e., even in the vicinity of 13~days)  are accurate.

\begin{figure}  [t]
\centering
 \includegraphics[width=8cm]{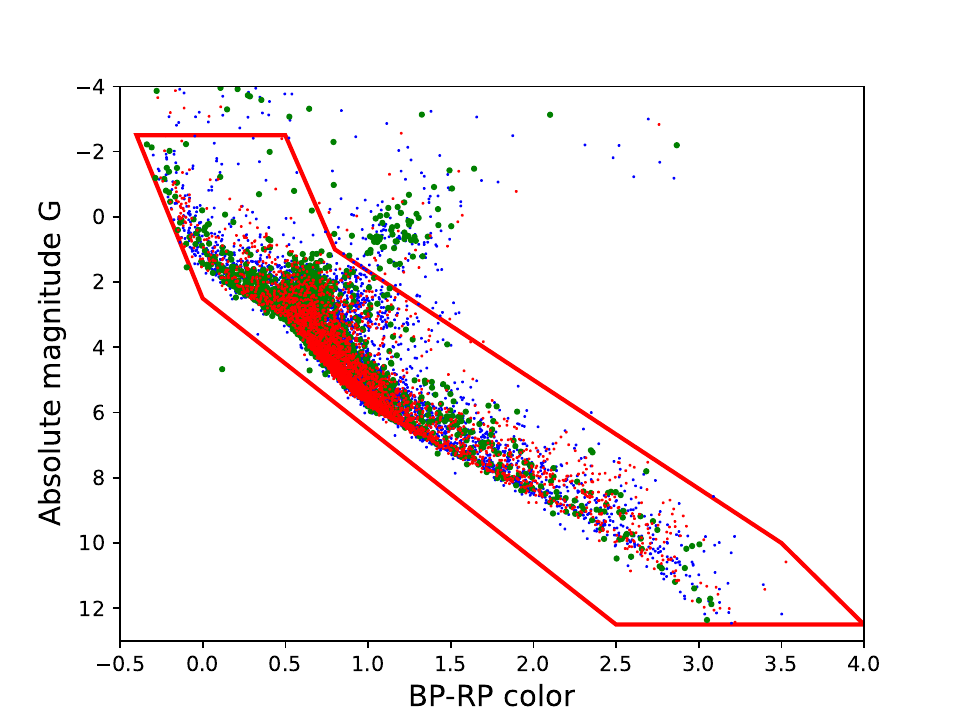}
 \caption{Color magnitude diagram (based on {\it Gaia} values) for eROSITA detected stars with 
 successful TESS period measurements, with the
indicated red polygonal region defining our MS sample.  The data points are
color coded according to period quality: green (grade 3), blue (grade 4), and red for higher grades. See text for details.}
\label{hrd_with_good_per}
\end{figure}

\section{Results and discussion}
\label{sec_result_disc}
\subsection{Construction of a stellar sample}
\label{stell_samp}
With our period measurements, we  proceeded to construct a well-defined stellar sample for all our further investigations.   To this end, we placed
the X-ray sources with successful period measurements into a the color-magnitude diagram.   Also, 
we had to take into account optical contamination.   As discussed by \cite{schmitt2022}, the eROSITA signals
for brighter objects can be corrupted by optical light and not be due to X-ray radiation.  To assess
the possible level of optically induced X-ray flux, we used the expression derived by  \cite{schmitt2022} to compute an
expected optical contamination rate, c$_{opt}$, and demand that the actually measured rate,
c$_X$, exceeds c$_{opt}$ by at least a factor of 5.

In Fig.~\ref{hrd_with_good_per}
we show a color-magnitude diagram of the selected sources, where we also indicate
a polygonal region, which we consider to contain bona fide MS objects; objects inside the polygon are indicated with the MS flag in Table~\ref{tab2}. 
It is obvious that the
vast majority of objects  with accepted period measurements are indeed MS stars (10798). Outside the
polygonal region in Fig.~\ref{hrd_with_good_per}, we find less than 2\% of the overall population.   The assigned period quality grades are
color-coded in Fig.~\ref{hrd_with_good_per}, the green data points are due to sources with the lowest grade (3) and
appear to be clustered for giant stars. To avoid possible problems with contamination, we only considered periods with grade 4 or higher
in the following.

\begin{figure}  [h]
\centering
 \includegraphics[width=8cm]{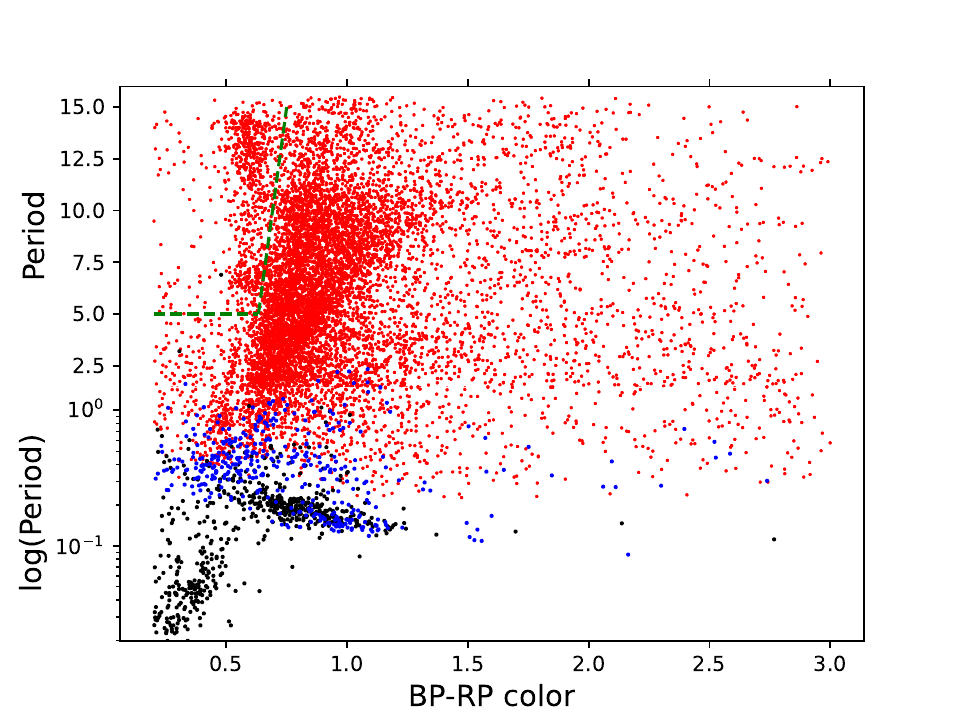}
   \caption{\label{colper} Period (in days) vs. BP-RP color for the MS sample with grade 4 and higher; black symbols refer to stars with 
   periods in excess of the estimated break-up period,  blue symbols refer to stars with more than 50\% of breakup, red symbols to those below. Note: the period scale for periods below 1.5~d is logarithmic.  
The periods for the stars inside the green dashed lines are not thought to be rotational periods. See text for details.
}
  \end{figure}
  
Next, we considered the relationship between the BP-RP colors and periods for those X-ray sources associated with MS objects
and successful period determinations; namely, the stars inside the red polygon in Fig.~\ref{hrd_with_good_per} and grade 4 and higher, which is shown
in Fig.~\ref{colper}.
As is apparent from Fig.~\ref{colper}, most of our sample stars are located in the BP-RP color range 0.5-1.0; namely, in the range of F and G type stars.
We first consider the short-period regime, where we notice two specific regions:
one region (1) with objects in the color range 0.0 $<$ BP-RP $<$ 0.5 and periods below 0.1~d with a tendency for
increasing period with increasing color, and a second region (2) of sources in the color range 0.5 $<$ BP-RP $<$ 1.2 and periods
between 0.3~d $>$ P $>$  0.1~d, with a tendency of decreasing periods with increasing colors.     
  
To interpret the short-period sources shown in Fig.~\ref{colper}, it is of interest to examine the break-up velocities and periods for 
the MS sample stars.   In this context we define the break-up
period, $P_{br}$, as 
\begin{equation} 
\label{def_per_break}
P_{br} = \frac {2 \pi R}{{\text{v}_{br}}}
,\end{equation}
with the break-up velocity, $\text{v}_{br}$, defined in the usual way as 
\begin{equation} 
\label{def_b_break}
{\text{v}_{br}} = \sqrt{\frac {G M}{R}},
\end{equation}
with $M$ and $R$ denoting stellar mass and radius, respectively, and G is the gravitational constant.  In Fig.~\ref{colper}, those stars
with measured periods of more than half the calculated break-up period are shown as blue and black data points and,  as expected, these stars
populate the short-period region in Fig.~\ref{colper}.   For an hypothetical binary consisting of two solar-like stars in contact, we can compute
a period of 0.23~d and P $\sim$ M. Thus, we expect that the stars in region (2) are close or contact binaries, whereas the stars in region (1)
are probably pulsating sources.   While we would expect identical orbital and rotation periods for close binaries, 
it should be clear that those periods indicated in Fig.~\ref{colper} with black data should not be treated as pure rotation periods. 

Finally we note a region where stars with bluish colors and periods in excess of 5~days are located, delineated by the
green dashed line in   Fig.~\ref{colper}. We can further note that there appears to be a clustering of periods near 6.5~days and 13~days.
We suspect that these periods are instrumental and ignore these data points inside the green dashed line in Fig.~\ref{colper} in our subsequent analysis.

\subsection{Some exemplary stars of low activity}
\label{sec_act_low}

Before discussing and interpreting our eROSITA/TESS results in detail,
we introduced four additional rather well known low-activity stars for purposes of comparison 
and checking the validity of the rotation-activity paradigm.
We specifically refer to the A-type stars Altair (= $\alpha$ Aql) and  Alderamin   (= $\alpha$ Cep),
the Sun, and Barnard's star; for these stars we have no eROSITA measurements, but X-ray flux and rotation 
period measurements are available from other sources.

Altair is a nearby late A-type star of spectral type A7IV-V, considered to be the ''hottest magnetically active stars in X-rays'' by
\cite{robrade2009}; these authors report on a long XMM-Newton pointing on Altair, carried out a detailed spectral analysis of the
X-ray data and find an X-ray luminosity of 1.4 $\times$ 10$^{27}$ erg/s
and a logarithmic L$_X$/L$_{bol}$ ratio of -7.4.   Altair has long been known to be a very rapid rotator with v sin(i) values exceeding 200 km/s;  
\cite{vanbelle2001} interferometrically resolved Altair and explicitly derive its oblateness due to its rapid rotation.  Altair's rotation period turns out to
be about 9.5 hours, and its equatorial rotation velocity amounts to a substantial fraction of its break-up velocity.   Based on TESS observations of Altair,
\cite{rieutord2024} derive an age of 88 Myr based on stellar oscillations, while earlier age determinations based on isochrones described
by \cite{lachaume1999} yielded ages of about 1.2 Gyrs.  Thus, Altair appears to be definitely younger than the Sun, maybe by a lot, it rotates about 65 times faster
than the Sun, yet it produces about the same X-ray output.  However, once the X-ray emission is scaled by L$_{bol}$, we find a logarithmic 
L$_X$/L$_{bol}$ ratio of -7.4,  much lower than that of the Sun.

Alderamin  is quite similar to Altair; it is also of spectral type A7IV-V, rotates very fast and again
its oblique shape could be measured interferometrically by \cite{vanbelle2006}.  The latter authors deduce a rotation period of 12 hours and 
conclude that Alderamin is rotating at 82\% of its break-up velocity.
\cite{huensch1998} find a rather weak X-ray source in the ROSAT all-sky survey data at the position of Alderamin with an X-ray count rate 
almost identical to that of Altair once scaled by the squared distance ratio, thus the X-ray luminosities are very similar.  With the bolometric luminosity derived by 
 \cite{vanbelle2006} we find a logarithmic  L$_X$/L$_{bol}$ ratio of -7.7.
Due to the optical brightness of these two stars no reliable {\it Gaia} data is available; the B-V colors of
both stars are identical, we thus assume the same for the BP-RP colors and use a value of  0.26.   

Our third comparison star is the Sun, for which we obviously do not have direct
X-ray measurements with ROSAT, XMM-Newton, eROSITA etc..  Determining the total solar X-ray luminosity from solar data is somewhat
cumbersome, and we refer to \cite{judge2003} for a detailed discussion of this issue.   For our purposes we assume a solar X-ray luminosity
of 2 $\times$ 10$^{27}$ erg/s, which ought to be a fair cycle average at soft X-ray wavelengths, however, as pointed out by \cite{judge2003}, 
the peak-to-peak
cycle variations can be up to a factor of ten even outside flares.   The equatorial rotation period of the Sun is
is well known to be 25.67~d, and we assume a  BP-RP color of 0.82.   With this X-ray luminosity, we obtain an  L$_X$/L$_{bol}$ ratio of -6.3.

Finally, we consider a very nearby star, Barnard's star, which is of very late spectral type M4 with a
BP-RP color of 2.83; \cite{hernandez2024} provide a detailed discussion of the rotation period of Barnard's star and conclude that it is about 150~d.
Previously, X-ray emission from Barnard's star had been reported by \cite{schmitt1995} at a level of about 4 $\times$ 10$^{25}$ erg/s, thus making 
it one of the weakest known X-ray sources outside of the solar system.   With a bolometric luminosity of 3.5 $\times$ 10$^{-3}$ L$_{\odot}$, 
we then finds a logarithmic  L$_X$/L$_{bol}$ ratio of -5.54, thus the star with
the lowest X-ray luminosity and slowest rotation is the most active one when considered w.r.t. L$_X$/L$_{bol}$.   It is very important to realize that the
X-ray luminosities of our sample stars with successful period measurements are very much different from these selected low activity stars as
demonstrated in Fig.~\ref{colper}, which re-emphasizes the point that a shallow all-survey as eROSITA misses out
the bulk of the low-activity stellar population with X-ray properties like our Sun, this population can only be detected -- if at all -- in the immediate vicinity
of the Sun and therefore contributes only very small numbers to the overall X-ray source population.

\begin{figure}[t]
\includegraphics[width=\linewidth]{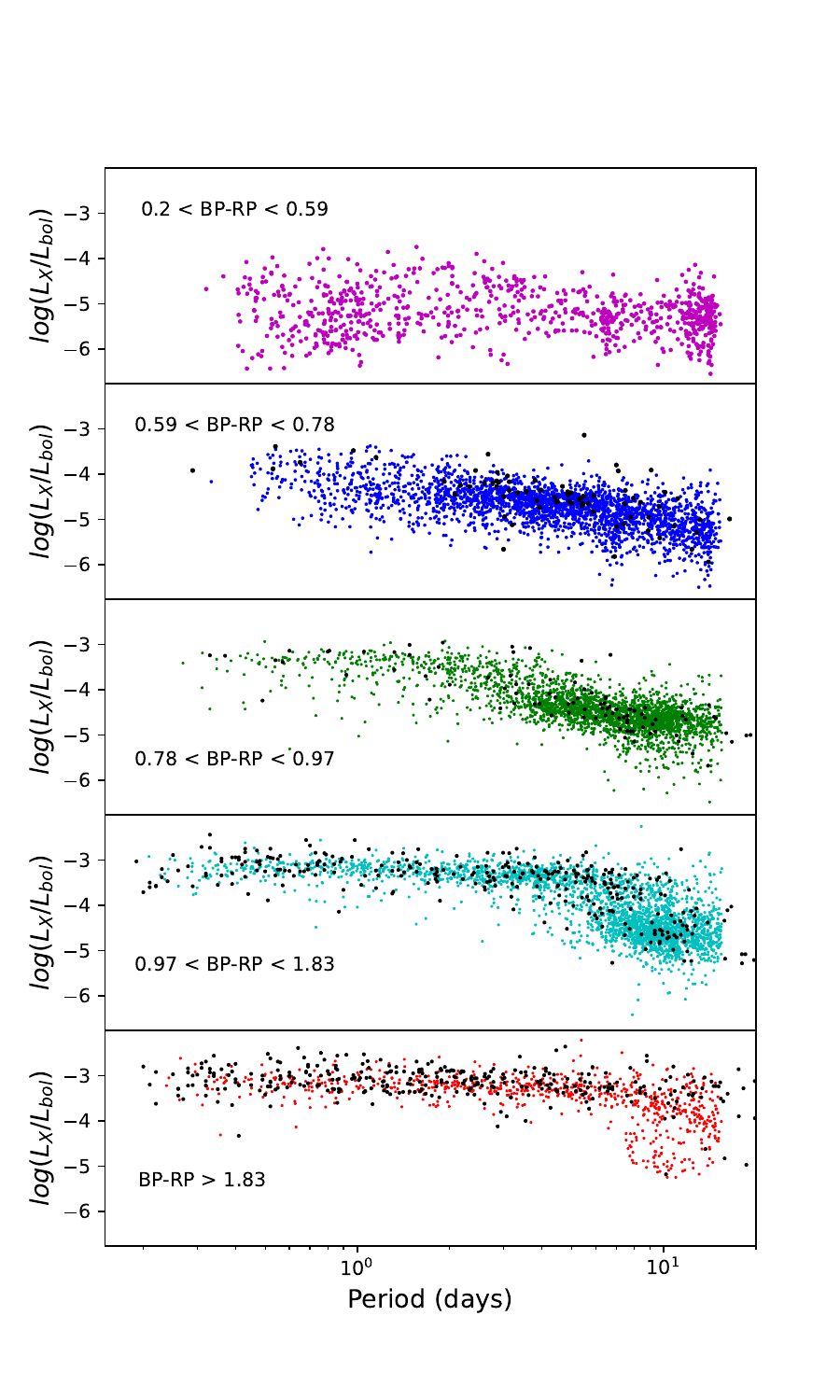}
\vspace{-1.cm} 
\caption{\label{lxlbolvsper}  log(L$_X$/ L$_{bol}$) vs. logarithmic period for MS stars in our sample;
 the stars studied by \cite{2011wright} are indicated by black dots. See text for details. 
 Top panel: Stars in the color range 0.2 $<$ BP-RP $<$ 0.59, dubbed early F-type stars.
 Second panel from top: Stars in the color range 0.59 $<$ BP-RP $<$ 0.78, dubbed late F-type stars. Third panel from top: Stars in the color range 0.78 $<$ BP-RP $<$ 0.97, dubbed G-type stars.
Second panel from bottom: Stars in the color range 0.97 $<$ BP-RP $<$ 1.83, dubbed K-type stars. Bottom panel from top: BP-RP $>$ 0.78, M-type stars.}
\end{figure}

\subsection{Period activity relations}
\label{sec_per_col}

We  return to a discussion of  Fig.~\ref{colper}, where we can very clearly notice the so-called onset of convection, 
a phenomenon that is well known in X-ray studies
of late-type stars \citep{schmitt1985}. Here, the number of sources with colors BP-RP $>$ 0.6 and periods~$>$1~d is rapidly increasing.
However, we also notices two regions with somewhat ''strange'' rotation periods (i.e., in the color range  $\approx$~0.4~$<$BP-RP< $\approx$0.65
with rotation periods near 13 days) and another alias peak near 6 days.   It is likely that
the accumulation of stars near these periods is instrumental; therefore, we flag the sources inside the green dashed line in  Fig.~\ref{colper} to
be able to identify these sources in our subsequent analyses; these sources are marked with the flag PB in Table~\ref{tab2}.
Finally, we note that
the paucity of stars with colors in excess of BP-RP $\sim$  1.2 (as seen in Fig.~\ref{colper}) is a selection effect. This is because in a flux-limited X-ray survey, the sampled volumes for redder dwarf stars become smaller and smaller with increasing color,
since the X-ray luminosities of the underlying stars are limited by the so-called saturation limit, which we discuss
in detail in Sect.~\ref{sat_cool_stars}.  

\subsubsection{X-ray activity versus period}
\label{sec_act_per}

In this section, we study how the X-ray activity of our sample stars as measured through the fractional
X-ray luminosity, L$_X$/ L$_{bol}$, varies with the measured TESS periods. For comparison, we also add the stars
already studied by \cite{2011wright} using  the same X-ray data and period measurements available at the time, yet with their
V-K colors converted to BP-RP colors using the tables presented by \cite{2013pecaut}.
We consider stars of type F, G, K, and M separately, distinguish between early and late F-type stars and plot the measured
L$_X$/L$_{bol}$-ratios versus period in Fig.~\ref{lxlbolvsper}.

\begin{figure}  [thb]
\centering
\vspace{-0.5cm} 
\includegraphics[width=8cm]{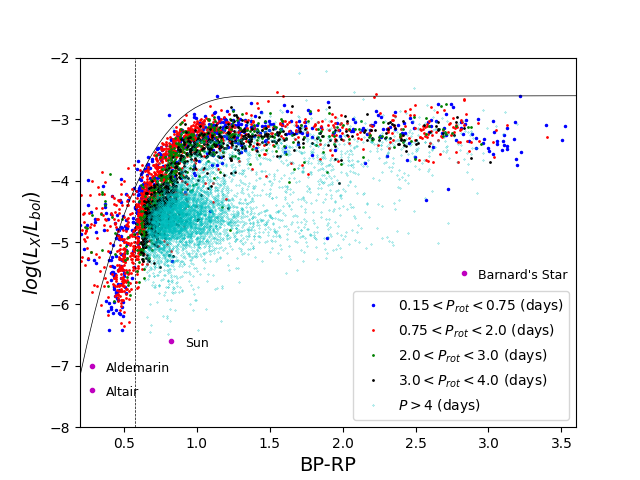}
\caption{\label{lxlbolvscolor} L$_X$/ L$_{bol}$ vs. BP-RP color for stars in various period ranges.    The stars studied by
\cite{2011wright} are indicated by black dots. See text for details.}
 \end{figure}

Starting with the  M-type stars displayed in the bottom panel of Fig.~ \ref{lxlbolvsper}, we note that
the short-period stars in this spectral range do reach a level of about L$_X$/ L$_{bol}$~$\sim$~10$^{-3}$, which is maintained for rotation 
periods of up to about seven days, and for longer periods, we see a clear drop in the activity levels;  
as also evidenced by Fig.~ \ref{lxlbolvsper}, this spectral range is well populated by stars contained in the sample by  \cite{2011wright}.  
Moving on to the K-type stars in our sample (displayed in the second panel from the bottom in Fig.~ \ref{lxlbolvsper}), we again
find that the short-period stars in this spectral
range do reach a level of about L$_X$/ L$_{bol}$ $\sim$ 10$^{-3}$, a bit lower than that reached by the M-type stars;
this level is maintained for rotation periods of up to about
four days and for longer periods, we see a clear drop in the activity levels.  This spectral range is also well populated by
stars contained in the sample by  \cite{2011wright}.  Considering next the G-type stars in the medium panel of  Fig.~ \ref{lxlbolvsper},
we clearly note that the saturation limit for G-stars does not reach  
the value of L$_X$/ L$_{bol}$~$\sim$10$^{-3}$; rather, it is found at a value
of about L$_X$/ L$_{bol}$~$\sim$~6~10$^{-4}$ and a clear drop in 
L$_X$/ L$_{bol}$ is very much apparent apparent for periods longer than about 3 days. Hence, we observe a very
clear dependence of L$_X$/ L$_{bol}$ on the period.  The number of G-type stars in the short-period range reported
by \cite{2011wright} is rather small, yet the objects we are obviously missing in our eROSITA sample are truly low-activity 
stars, such as the Sun, with a period of 25~days or more and L$_X$/ L$_{bol}$-value of a few times 10$^{-7}$ (cf., Fig.~\ref{lxlbolvscolor}).
Finally, we turn to the F-type stars plotted in the two upper panels of Fig.~\ref{lxlbolvsper}, where we subdivided our sample in
early and late F-type stars using BP-RP = 0.59 as the break point between the two groups.
First,  it is clear that among the F-type stars, we found no star reaching the canonical saturation limit 
at L$_X$/ L$_{bol}$ $\sim$ 10$^{-3}$,
the empirical saturation limit for those stars with BP-RP $>$ 0.59 appears to lie closer to L$_X$/ L$_{bol}$~$\sim$10$^{-4}$; for the earlier F-types, it is near L$_X$/ L$_{bol}$~$\sim$10$^{-5}$ (if we want to talk about saturation at all).
We note  that these two groups of stars show different period dependences in their L$_X$/ L$_{bol}$-ratios: the earlier types (top panel in Fig.~\ref{lxlbolvsper})
do not show any period dependence whatsoever, while the later types (second panel from top in Fig.~\ref{lxlbolvsper})
appear to be constant for periods below about 2~days, followed by clearly decreasing
L$_X$/ L$_{bol}$-ratios, with increasing periods as seen for the stars of later spectral types.   

\subsubsection{X-ray activity versus color}
\label{sec_act_col}

For almost all spectral types of stars shown in Fig.~\ref{lxlbolvsper}, we  notice a saturation level  for
the shorter periods,  which, however, depends on spectral type.   While for the later type K and M type stars this level
lies a little bit below L$_X$/L$_{bol}$ $\approx$ 10$^{-3}$, such values are definitely not reached by F-type stars.
Furthermore, the activity-period dependence for stars below the saturation limit 
varies with spectral type:  even though we see that for the earliest F-type stars, no period dependence is
recognizable, the period dependence starts at around two days for the later types and that threshold moves up to around nine days for
M-type stars.   This finding is clearly not new, but was already realized in the studies by \cite{2003pizzo} and \cite{2011wright},
except that these studies contained only few G-stars and even fewer F-type stars (cf., Fig.~\ref{lxlbolvsper}).

With our new period measurements, we can now put the saturation limit in context and
revisit the relation between X-ray activity and color, differentiated according to the stellar period.   In Fig.~\ref{lxlbolvscolor}, we consider our sample
stars with secure periods and plot their activity as measured through their L$_X$/L$_{bol}$ ratio  as a function of BP-RP color with the
different period ranges indicated by different colors as indicated  in the legend of  Fig.~\ref{lxlbolvscolor}. For the purposes of orientation and discussion, we also added the four low-activity stars 
discussed in detail in Sect.~\ref{sec_act_low} in Fig.~\ref{lxlbolvscolor}.

\begin{figure}
\includegraphics[width=\linewidth]{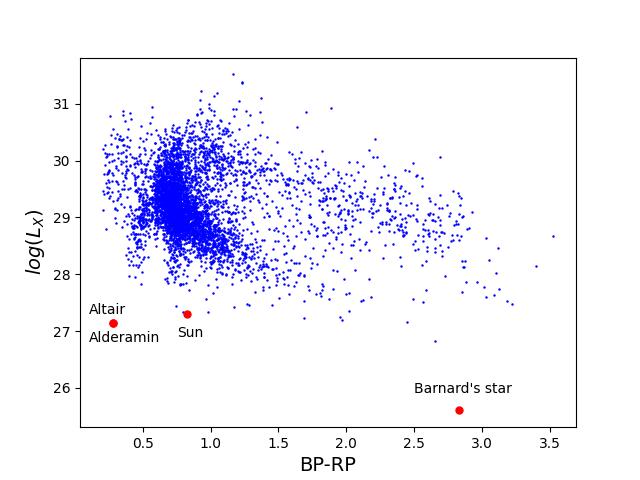}
\vspace{-0.1cm} 
\caption{\label{lxbprp}  X-ray luminosity L$_X$ vs. BP-RP color for sample stars with successful period measurements 
as well as the selected low activity stars (red data points). See text for details. }
\end{figure}

Figure~\ref{lxlbolvscolor} shows a very rather sharp upper boundary of the measured L$_X$/L$_{bol}$-ratios already apparent 
in the eROSITA population studies by \cite{freund2024},  however, as apparent from Fig.~\ref{lxlbolvscolor},
the most active stars in terms of L$_X$/L$_{bol}$ at a given BP-RP color are those with the shortest
periods, as expected. Furthermore, most of the ''noise'' apparent in Fig.~\ref{colper} has disappeared, a few stars remain in the spectral range
0.2~$<$BP-RP$~<$0.6, which  show no period dependence in their X-ray activity;  population simulations (cf., Sect.~\ref{sim_pop}) suggest
that these objects could be binaries with the low mass companion responsible for the observed X-ray emission. 
In  Fig.~\ref{lxbprp} we then examine the X-ray luminosities
of our sample stars; we specifically note that the stars in the color range 0.2~$<$BP-RP$~<$0.6  do not appear to differ substantially 
from those of later spectral types, which suggests a binary interpretation:  
these stars ought to be rather young, thus, if accompanied
by a late-type companion of type G or K,  X-ray luminosities in the range of a few 10$^{29}$ erg/s are not at all unusual; this would explain
both the lack of any dependence on rotation and the measured  L$_X$/L$_{bol}$-ratios.

Therefore, Fig.~\ref{lxlbolvscolor} demonstrates that activity of cool stars as measured through the L$_X$/L$_{bol}$ ratio
drops by two orders of magnitude over a rather narrow range of stellar color between 0.6~$<$BP-RP$<$1.0 corresponding to a spectral
type between about F6 and G9.   This could arguably have to do with some ''overcorrection'' of the X-ray luminosities by the bolometric
luminosities; however, the latter change only by a factor of six or so over the same spectral range, so this cannot be the correct
explanation.   Also, the rotation period alone cannot lead us to the correct explanation.   While the Sun and (even more so) Barnard's star are located
well below the observed L$_X$/L$_{bol}$-ratios encountered in the relevant color ranges,  the extremely fast rotators Altair and Alderamin
have the lowest activity values (in terms of  L$_X$/L$_{bol}$) of all stars considered in seeming contradiction to any rotation-activity paradigm.

\subsection{Convective turnover time and Rossby number}
\label{sec_tau_conv}

As made apparent in Sect.~\ref{sec_act_col},
we clearly have to look out for yet another stellar property to properly understand and interpret Fig.~\ref{lxlbolvscolor}. However, we must also consider 
what stellar property  changes by several orders of magnitude in the relevant spectral range.  The natural answer obviously is the 
property of the stellar convection zones and their dramatic changes, with K stars having rather deep convection zones 
comprising significant fractions of
stellar radius, while those of F-type stars become thinner and thinner.   However, we must also consider how this relates to the question of
activity saturation. 

As shown by \cite{noyes1984} for a sample of chromospherically active stars and by \cite{2003pizzo} and \cite{2011wright} 
for a sample of X-ray active stars, we can ``normalize'' the rotation periods of stars of different spectral type
by scaling the observed rotation periods with the so-called convective turnover time, which yields the Rossby number, 
and relate the activity properties
to this Rossby number:  by ''forcing'' the observed  distributions of  L$_X$/ L$_{bol}$ ratios and rotation periods, P$_{rot}$,
to a common scale for all their sample stars, 
both \cite{2003pizzo} and \cite{2011wright} 
were able to derive empirical convective turnover times (and eventually Rossby numbers) as a function of stellar color or stellar mass. 

Motivated by the good agreement between the convective turnover times as calculated by \cite{landin2023} and empirically derived ones
by \cite{2011wright},  we decided to proceed with an empirical fit to the $\tau _{conv}$ BP-RP relation.    
Following \cite{noyes1984}, we chose a third-order polynomial
for colors up to some value (BP-RP)$_{break}$, followed by a linear relation for redder colors.
 Abbreviating X = (BP-RP) - (BP-RP)$_{break}$, we can  construct the ansatz,

\begin{align}
\label{def_tau_conv}
\nonumber
log (\tau _{conv}(X))  &=  T_0 + C \times X  + B \times X^2  + A  \times X^3  \text{~~for~~~~~}X < 0,\\
                                      &=  T_0 + C \times X      \text{~~~~~~~~~~~~~~~~~~~~~~~~~~~~~~~~ for~~~~} X > 0,
\end{align}

\noindent
with four free parameters A, B, C, and $T_0$, which, together with the known BP-RP colors, determine $\tau _{conv}$ (specified in days).
This analytical ansatz also requires the specification of a break point between the two color regimes; we choose BP-RP = 1.2,
which agrees well with the computations by  \cite{landin2023} and is consistent with the choice of \cite{noyes1984}, who used B-V = 1.

Once the values of the convective turnover times are specified (e.g.,  those calculated from Eq.~\ref{def_tau_conv} 
using the measured BP-RP colors), we can compute a Rossby number
using the also measured period for every sample star and thence -- following the nomenclature by  \cite{2011wright}  -- produce a 
R$_X$ (= L$_X$/L$_{bol}$) versus Rossby number (Ro) diagram.  
Both  \cite{2003pizzo} and \cite{2011wright} used a (logarithmic) broken power-law description, namely, based on relationships between R$_X$ and 
Ro taking the form of

\begin{align}
\label{def_act_rossby_1}
\nonumber
R_X(Ro)  &= R_{X,sat} \text{~~~~~~~~~~~~~~~~~~~~~~~~~~~~for~~~} Ro < Ro_{break}, \\
                &=  R_{X,sat} \times \left({\frac{Ro_{break}}{Ro}}\right)^{\beta}  \text{~~~~~~~for~~~} Ro > Ro_{break},
\end{align}

\noindent
with the slope, $\beta$, the saturated Rossby number, R$_{X,sat}$, and the Rossby number break point, Ro$_{break}$, at the break point between the two dependences. This allowed them to devise a simple model, referred to as model A in the following,
with only three independent parameters.  Using this approach, \cite{2003pizzo} and \cite{2011wright} were 
able to obtain good empirical descriptions of the activity properties of their respective stellar samples.
Naturally, Eq.~\ref{def_act_rossby_1} is quite arbitrary and it might be preferable to adopt other forms without a break. One such possibility is given by the ansatz,

\begin{align}
\label{def_act_rossby_2}
R_X(Ro)  &=& A_0 \times e^{-Ro/\lambda} \text{~~~~~~~+~~~~~~~} A_{\infty} \times \frac{e^{-\lambda/Ro}}{Ro^{\beta}},
\end{align}
\noindent
with the free parameters A$_0$, A$_{\infty}$, $\lambda$, and $\beta$, referred to as model B in the following.
In Eq.~\ref{def_act_rossby_2}, we obviously have $\lim_{Ro \to 0} R_X(Ro) = A_0$ , where 
$A_0$ is the saturation limit and asymptotically for $\lim_{Ro \to \infty} R_X(Ro), $ we have R$_X$  $\sim$  A$_{\infty}$ Ro$^{-\beta}$ (i.e., the $\beta$ parameters 
in Eqs.~\ref{def_act_rossby_1} and~\ref{def_act_rossby_2} are the same).   Obviously, Eq.~\ref{def_act_rossby_2} is as arbitrary as Eq.~\ref{def_act_rossby_1} 
in the sense that is a purely empirical ansatz without any theoretical foundation, yet it yields a differentiable function. We note that  in Eq.~\ref{def_act_rossby_2}, the saturation 
level is reached only asymptotically.

\subsection{X-ray activity versus Rossby number}  
\label{sec_xray_ross}
 
To study the relationship between X-ray activity (as measured through the L$_X$/L$_{bol}$ ratio)
and  Rossby number, we used Eqs.~\ref{def_act_rossby_1} and~\ref{def_act_rossby_2}
as empirical descriptions for the computation of the expected activity (in terms of L$_X$/L$_{bol}$).  
However, instead of using binned versions of convective turnover time as \cite{2011wright} , we use the
analytical description of convective turnover time given by Eq.~\ref{def_tau_conv}.    Our model thus consists of the
three parameters, $\beta$, the saturated, $R_{X,sat}$,  and the Rossby number, $Ro_{break}$, for the model defined by Eq.~\ref{def_act_rossby_1}
and the four parameters  $A_0$, $A_{\infty}$, $\lambda$, and $\beta$ for the model defined by Eq.~\ref{def_act_rossby_2}, 
describing the X-ray activity-Rossby number relationship and of the  four  parameters A, B, C, and $T_0$, describing the
dependence of convective turnover time on BP-RP color. Thus, the overall models have seven or eight free parameters, respectively.
However, to reduce the number of free parameters we fix the value of $\beta$ at 2 (as in \citealt{2011wright}) and, next,
we chose the parameter  $Ro_{break}$ (in Eq.~\ref{def_act_rossby_1}) and $A_{\infty}$ (in Eq.~\ref{def_act_rossby_2}).
Furthermore, as discussed in Sect.~\ref{sec_tau_conv} the break point between the two regimes in the $\tau_{conv}$-BP-RP relation
is fixed at BP-RP = 1.2, which results in convective turnover times for the
Sun that agree with the usual expectations.

Given set of values for the parameters  A, B, C, and $T_0$, we can
compute the values of convective turnover time for each sample data point. Combined with the measured period, this
allows for an estimate of the Rossby number, $Ro$, for every sample star star and, hence, an estimate of
this star's $R_X$-value using Eq.~\ref{def_act_rossby_1} or Eq.~\ref{def_act_rossby_2}, which can be compared to the observed 
$R_X$ ratios.   While this appears to be a straightforward non-linear fitting problem with seven or eight parameters, we ran into a number of
difficulties in arriving at stable solutions.   The main problem here are interlopers (i.e., obviously wrong data data points).
Such interlopers could be caused by erroneous identifications of the X-ray sources, by incorrect period determinations,
by errors in the assumed stellar parameters, and (perhaps most importantly) through the presence of unrecognized
binaries. An illustrative example for the latter case is the well known nearby star $\alpha$ CrB, consisting of an A0V primary and
a G2V secondary component.  Fortuitously, this system is eclipsing and during optical secondary eclipse (i.e., A star in front of
G star) the system's X-ray flux drops to zero as shown by \cite{schmitt1993}, demonstrating that the X-ray emission exclusively 
comes from the later type star.
If this system were not known to be a binary system, our scheme would attribute the X-ray emission to a relative fast rotating 
A-type star, which would be clearly wrong, and such wrong data points can severely ''contaminate'' any straightforward ''best-fit'' solution.

To deal with such an interloper population  we  analyzed our data using a so-called ''mixture model,''
namely, the data were modeled by two populations, one genuine population obeying Eq.~\ref{def_act_rossby_1} or Eq.~\ref{def_act_rossby_2} and 
an interloper population modeled by 
a noise distribution.    Every data point was assigned a probability, $q$, of being part of the genuine population and $1-q$ to be part of the
interloper population. For a detailed explanation and discussion of this approach, we refer to the illuminating blog by  
\cite{foreman2014}, who also provides {\it Python} code examples.   

We  set up a mixture model composed of two populations: one
genuine population described by the
seven or eight parameters, A, B, C, $T_0$, and $\beta$, R$_{X,sat}$  and Ro $_{break}$ or $A_0$, $A_{\infty}$, $\lambda,$ and $\beta,$ along with some fraction, $q$;
and an interloper population, following \cite{foreman2014} described by a Gaussian distribution with some mean and dispersion, contributing
a fraction of $1-q $ to the overall population.  To solve this problem, we used the emcee code developed by \cite{foreman2013}, which utilizes
an affine-invariant ensemble sampler for a Markov chain Monte Carlo (MCMC) computation, as introduced by \cite{goodman2010}.   We iterated
the emcee produced MCMC chains until convergence, which we considered to be achieved when the total likelihood no longer showed any trends and, 
thus, the MCMC burn-in phase was achieved.   From this state, we start the production run for which we use 64 walkers and 1000 iterations.
For both the initial burn-in and subsequent production, we would need to put priors on some parameters.   As in \cite{wright2011}, we needs to constrain
the slope of the rotation-activity relation (i.e., setting the $\beta$ parameter in model A to 2) 
to arrive at stable solutions. In model B, we fixed the $\beta$ parameter again at 2 and the $A_0$ parameter at 3.   The priors for the remaining
parameters were chosen as flat priors with generous ranges around the ''expected'' values; care must be applied to the prior for the
saturation limit and the $\lambda$ parameter (in case of model B) to ensure that the MCMC chain converges at a reasonable solution.

\begin{figure}  [th]
\centering
 \includegraphics[width=8cm]{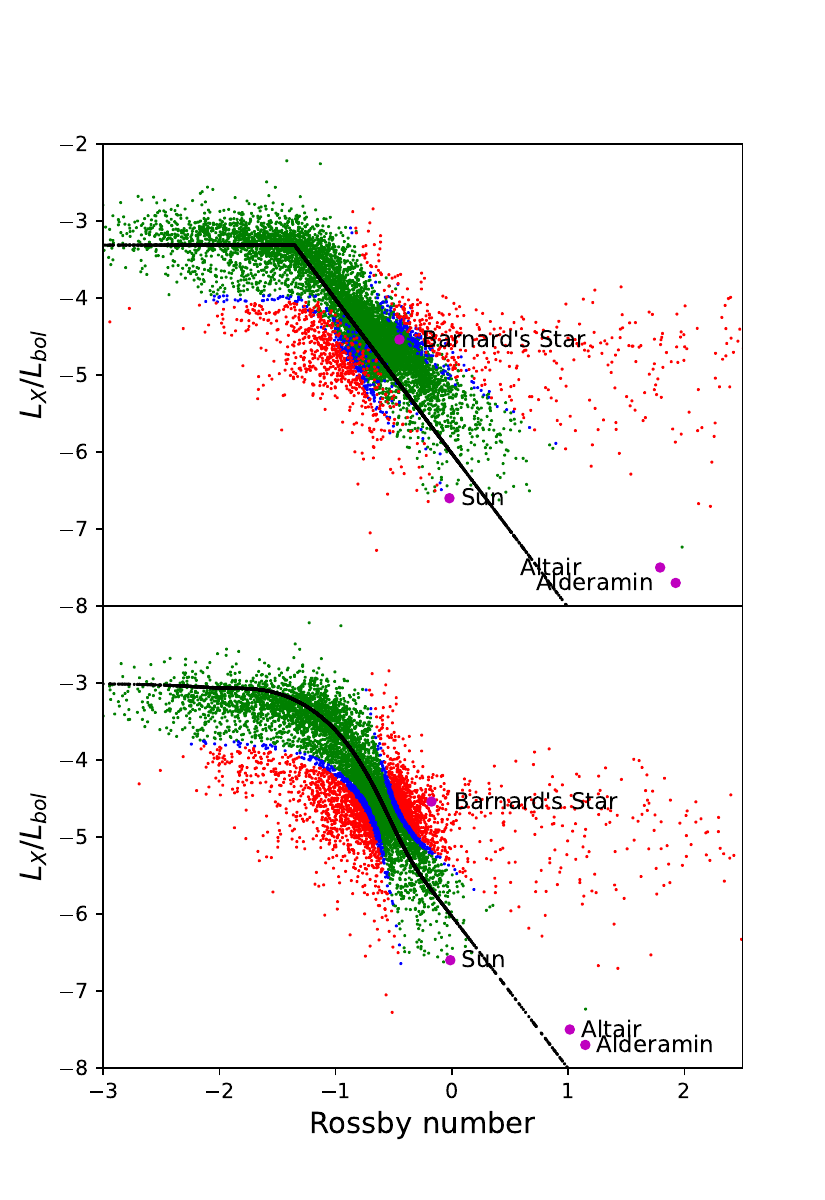}
 \vspace{-0.15cm} 
   \caption{   \label{lxlbolros} L$_X$/L$_{bol}$ ratio  vs. Rossby number for our sample stars with the different colors indicating different
   membership probabilities (green largest, red lowest), the black line is the best fit with the functional form as 
 given by Eqs.~\ref{def_act_rossby_1} (top panel)
   and~\ref{def_act_rossby_2} (bottom panel). See text for details. }
  \end{figure}

\begin{figure}  [th]
\centering
 \includegraphics[width=8cm]{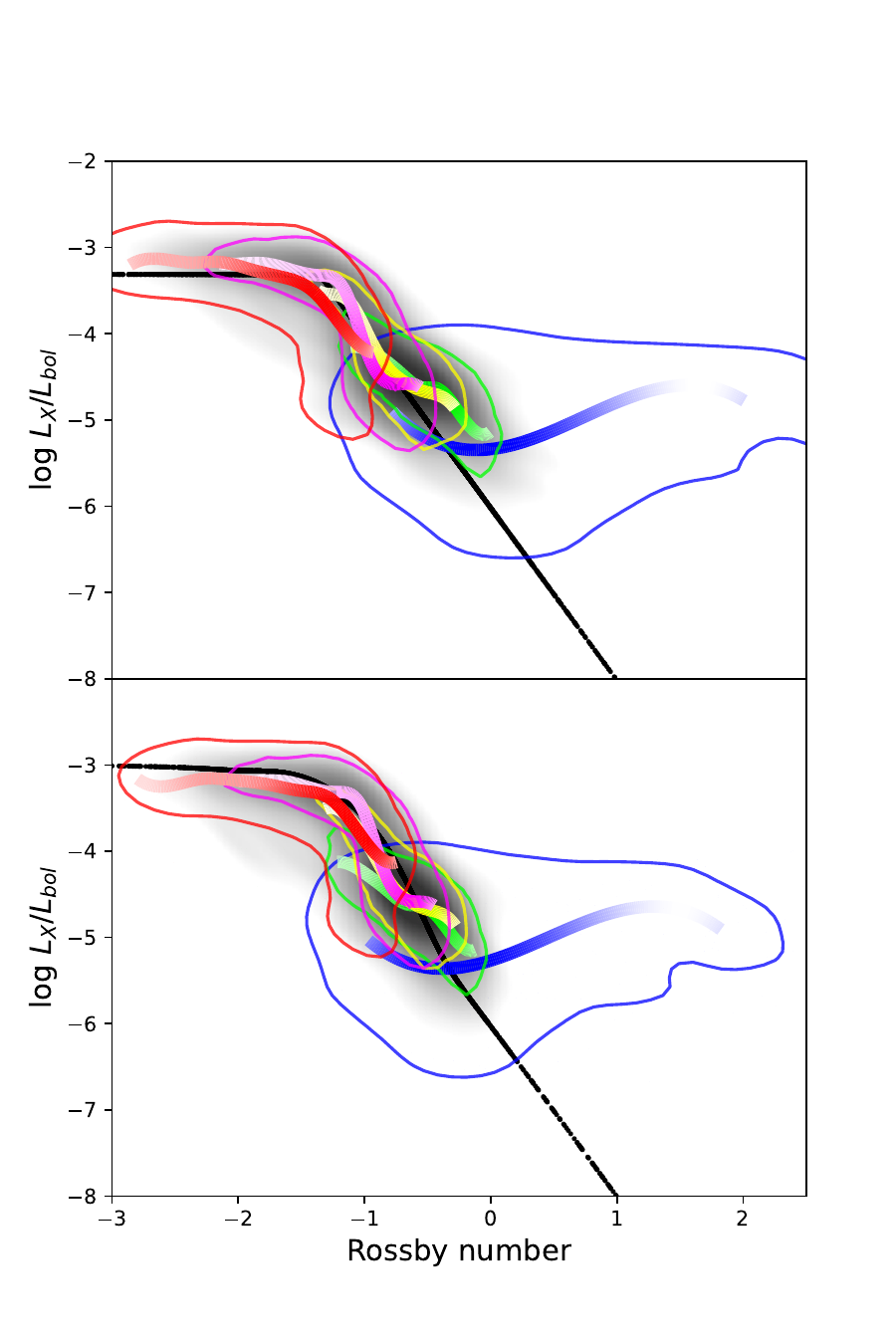}
 \vspace{-0.1cm} 
   \caption{\label{lxlbolros_sample} $L_X/L_{bol}$ ratio  vs. Rossby number for our sample stars with the different colors indicating 
   the different spectral types:  blue indicating A and early F-type stars, green late F-type stars, yellow G-type stars, magenta K-type stars, and
   red M-type stars. Population medians (thick line) and the 90\% contours (thin lines) of the population are also shown.
   The gray-shaded area indicates the density of the data points,  the black line is the best fit with the functional form as 
 given by Eqs.~\ref{def_act_rossby_1} (top panel)   and \ref{def_act_rossby_2} (bottom panel). See text for details. }
  \end{figure}

From the production run chains,
we computed the mean values and dispersion of all fit parameters and in particular for the
parameters A, B, C, and $T_0$, which define the convective turnover times.   Following the approach described by  \cite{foreman2014},
we then computed membership  probabilities for each individual sample star.
In Table~\ref{tab3} and in  Figs.~\ref{lxlbolros} and~\ref{lxlbolros_sample}, we show the results of these computations.   In Table~\ref{tab3}, 
we present the derived fit parameters for the two cases considered.  Each of the figures contains two panels,
showing L$_X$/l$_{bol}$ versus Ro with the top panels referring  to the prescription   Eqs.~\ref{def_tau_conv} and~\ref{def_act_rossby_1},
while the bottom panels use the prescription  Eqs.~\ref{def_tau_conv} and~\ref{def_act_rossby_2}.
In Fig.~\ref{lxlbolros} we concentrate on the individual data points, while in Fig.~\ref{lxlbolros_sample} we concentrate on the sample properties
with respect to spectral type.   In  Fig.~\ref{lxlbolros} the membership probabilities, $p_{mem}$, of our sample stars are color-coded; green data points indicate
stars with  $p_{mem} > 0.5$, blue data points stars with $0.3 < p_{mem}  < 0.5$, and red data points indicate stars with $p_{mem} < 0.3$.
Also shown in Fig.~\ref{lxlbolvscolor} are the four stars Altair, Alderamin, the Sun, and 
Barnard's star discussed in detail in Sect.~\ref{sec_act_low};   while Barnard's star lies in the range
of sampled L$_X$/L$_{bol}$ ratios, the Sun and in particular Altair and Alderamin lie several orders of magnitude
below the sampled activity range, but within the reach of the X-ray activity versus Rossby number 
relation defined by Eq.~\ref{def_act_rossby_1} or Eq.~\ref{def_act_rossby_2}, even if the scatter around the median is 
quite substantial.  

The many partially overlapping data points in Fig.~\ref{lxlbolros} obstruct our view of the sample properties.   Therefore, 
in Fig.~\ref{lxlbolros_sample} we concentrate on the sample properties differentiated by spectral type.
The overall density of our data points is
indicated by the gray shaded area, the medians of the distributions are given by the blue (A-type stars), green (F-type stars),  yellow (G-type stars),
magenta (K-type stars) and red (M-type stars) thick lines, the contours are also shown.   
The expected mean relations between the L$_X$/L$_{bol}$ ratio and Rossby number
are provided by the black colored line, calculated with the parameters specified in Table~\ref{tab3}.
A comparison of Figs.~\ref{lxlbolros} and~\ref{lxlbolros_sample} shows that the forbidden red data points in Fig.~\ref{lxlbolros} exclusively correspond
to A-type stars and therefore very likely do not indicate emission from these objects in contrast to the cases of Altair and Aldemarin shown in Fig.~\ref{lxlbolros}.
The medians of the other stellar types agree rather well although there might be differences in particular for the F-type stars; as noted earlier, the F-type and 
G-type stars do not reach the saturation level of $\approx$ 10$^{-3}$ in L$_X$/L$_{bol}$ and never reach Rossby numbers of 0.01.

Some support for this scenario can be derived from {\it Gaia} data.  We specifically investigated 160 large Rossby number sources shown in Fig.~\ref{lxlbolros}, i.e.,
source with formal Rossby numbers in excess of 10.  Out of these 16 (i.e., 10\%) were found to have the non-single-star flag set in the {\it Gaia} catalog  implying that
the binary fraction in this subsample appears to be much larger than in the {\it Gaia} catalog at large.  However, a systematic investigation of this issue
would go far beyond the scope of this paper.

\begin{center}
\begin{table}
\caption{\label{tab3} Fit results convective turnover time and Rossby number.} 
\begin{tabular}{ r r r r }\\
\hline
& ER.~\ref{def_tau_conv} $+$ Eq.~\ref{def_act_rossby_1} & & Eq.~\ref{def_tau_conv} $+$ Eq.~\ref{def_act_rossby_2} \\
\hline
A             & 5.11 $\pm$ 0.15 &                  A       & 4.74 $\pm$ 0.154 \\
B             & 1.18 $\pm$ 0.10 &                  B      & 1.15 $\pm$ 0.10  \\
C             & 0.506 $\pm$ 0.006          &              C      & 0.42 $\pm$ 0.01 \\ 
T$_0$               & 18.30 $\pm$ 0.005 &            T$_0$ &16.722 $\pm$  0.005 \\
log(R$_{X,sat})$      & -3.313 $\pm$ 0.008 &  A$_0$ & -3.0 (fixed) \\
                         &                                  &$\lambda$ & 6 $\times$ 10$^{-2}$  $\pm$ 8 $\times$ 10$^{-5}$\\
$\beta$     & 2.0 (fixed)                        & $\beta$     & 2.0 (fixed)\\
\hline
\end{tabular}                   
\end{table}
\end{center}

\begin{figure}  [bht]
\centering
 \includegraphics[width=8cm]{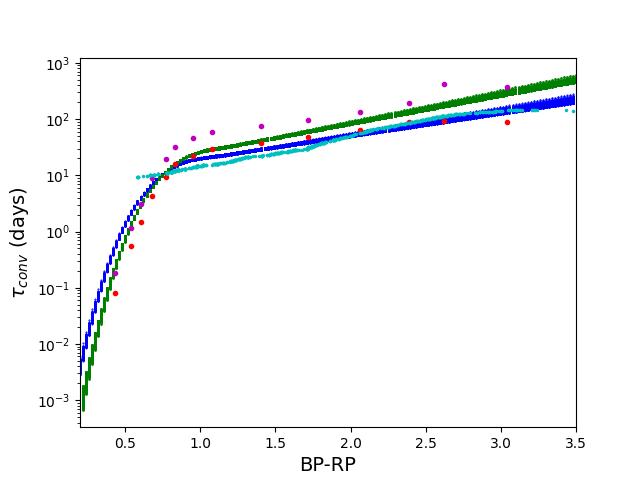}
  \vspace{-0.1cm} 
  \caption{   \label{tauconvnew} "Best-fit" convective turnover time vs. BP-RP colors (derived from Eq.~\ref{def_act_rossby_1} as 
  blue solid curve, derived from Eq.~\ref{def_act_rossby_2} as 
  green solid curve);  the shown data points (global convective turnover time in red, local convective turnover time in magenta)
  come from the theoretical calculations  by \cite{landin2023}, the cyan data points are derived from the sample presented by 
  \cite{wright2011}. See text for details.}
  \end{figure}

\subsection{Convective turnover time}

As a byproduct of our fitting exercise, we also obtained the parameters A, B, C, and $T_0$, which define our estimate
of the convective turnover time as a function of BP-RP color.   The results for the nominal fit parameters are shown
in Fig.~\ref{tauconvnew}; from the MCMC chains, we estimated the parameters  A, B, C, $T_0$, and their dispersion,
which define the estimates for the convective turnover times for the case of Eqs.~\ref{def_act_rossby_1} (blue curve) and 
\ref{def_act_rossby_2} (red curve).   For comparison, we also show the results of the model calculations by \cite{landin2023}
as well as the data points provided by \cite{wright2011}.
Our best-fit curves follow  the values for global and local turnover time for most of the parameter range rather closely;  however, for smaller BP-RP colors,
we obtained slightly larger values.   Also, we remark that this
value only sets the scale for the Rossby number.   We  further note that the values computed by  \cite{2011wright} reproduce the
trend for colors BP-RP $>$ 0.9, but obviously the range of F-type stars and early G-type stars is not really covered by the sample
of \cite{2011wright}  and the massive decrease in convective turnover time in this spectral range is missed altogether.

\begin{figure}  [bht]
\centering
 \includegraphics[width=8cm]{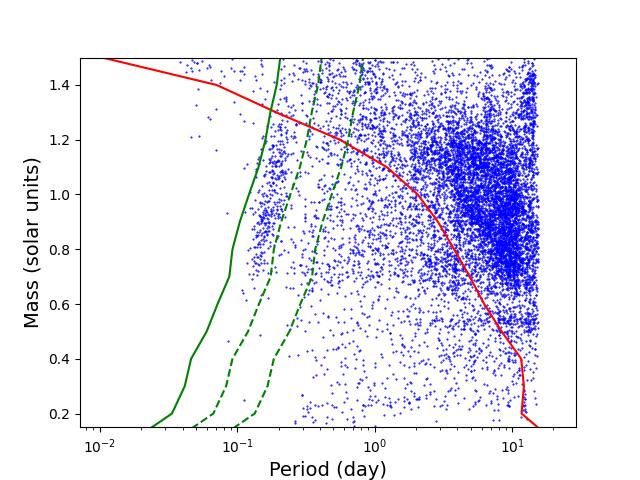}
  \vspace{-0.cm} 
  \caption{   \label{massper} Mass (in solar units) vs. period (in days) for our MS sample, with masses estimated from the
 stars' BP-RP colors using data tables from \cite{2013pecaut}.  The red solid line is the period computed from Eq.~\ref{def_act_rossby_1} at the
 Rossby break point, $Ro_{break}$, and the convective turnover time computed from Eq.~\ref{def_tau_conv}. The green lines show the 
 break up periods computed from  Eq.~\ref{def_per_break}. See text for details.}
  \end{figure}
  
\subsection{The saturation limit and cool stars}
\label{sat_cool_stars}

Clearly, this leads  us to wonder whether the agreement between theory and observations is coincidental or physical.  However, if the relations given by 
Eqs.~\ref{def_act_rossby_1} or~\ref{def_act_rossby_2} and
the run of convective turnover time with mass (and, hence, BP-RP color) hold over the whole parameter
range of cool stars, then it is clear that stars with increasing mass (and, 
thus, decreasing convective turnover time) find it exceedingly difficult to reach the small Rossby number regime
and, hence, the saturation limit, since they would have to rotate with periods substantially below their 
convective turnover time, which are becoming quite small already.
Using the expression in Eq.~\ref{def_tau_conv} and our fit results, we can address the question of which period range 
is available to stars with respect to a rotation that is, on the one hand,  
slow enough to stay above their respective break-up period, but rotate, on the other hand,  
rapidly enough to enter the saturation regime.    The results are shown in 
 Fig.~\ref{massper}, where we plot the measured periods and estimated masses for our sample stars.
 Also shown is the convective turnover time (red line, as computed from Eq.~\ref{def_tau_conv}) 
 as well as the estimated break-up periods for 100\%, 50\%,
and 10\% of the estimated break-up level (green lines).    
To enter the saturation regime without breaking up, a star must lie below the red and green curves.
As is obvious from Fig.~\ref{massper}, there is only a restricted range of periods available for 
saturation and stars with masses above $\approx$ 1.2 M$_{\odot}$   simply cannot rotate rapidly 
enough to ever enter the saturation regime. 

The problem of X-ray saturation has been around for a while and \cite{2011wright} presented a detailed account of this issue.
\cite{2011wright} specifically stated that in their sample "a mean saturation level of log(R$_X$) =  -3.13 $\pm$ 0.22, almost independent
of spectral type'' was  found, yet they also stated that "a significantly lower mean saturation level for the highest-mass stars in their sample
(F-type stars)" was found as well.  However, \cite{2011wright} then argued that these stars (i.e., F-type stars) are not saturated 
but ''super-saturated'' and that therefore the saturation level is independent of spectral type.  We note that super-saturation refers to the 
phenomenon that for very fast rotators, the fractional
X-ray luminosity, R$_X$, appears to decrease, yet the evidence for the phenomenon is scarce as stated by \cite{2011wright}.  
Furthermore, these authors discuss in particular
the so-called ''polar updraft'' effect introduced by \cite{stepien2001} in a study of W~UMa-systems and the so-called ''centrifugal stripping'' effect, introduced by
\cite{james2000} in a study of M dwarfs.    In the case of a rotating star, the von Zeipel theorem states that
the emergent radiative flux is proportional to the local
gravity, which becomes smaller and smaller with increasing rotation rate, and as a consequence the flows and hence convective motions in the polar direction
differ substantially from those in equatorial direction.   In the ''coronal stripping'' scenario developed by \cite{james2000},  the location
of the co-rotation radius with respect to the stellar radius is a primary focus.    Inspecting the relevant formulae for both scenarios, we realize right away that the critical point is reached if a
star were to rotate with its break-up period given by Eq.~\ref{def_per_break}; in this case, the equatorial effective gravity would vanish and the co-rotation radius would coincide with the stellar radius.

For the more massive stars, these scenarios yield almost identical results, since the radius of the radiative-convective boundary moves close to the stellar radius and at the equator the two expressions (Eqs.~6 and~8 in \citealt{2011wright}) become identical.  Furthermore, R$_{Kepler}/R_*$ = 1 and $G_X = -1$ (in the notation
of \citealt{2011wright}) represent the absolute limit if we postulate  that stars cannot rotate beyond break-up, with the consequence that most of the region dubbed ''supersaturated'' in Fig.~5 of \cite{2011wright} cannot be populated at all.   
Evaluating then the break-up period of a star with a mass of 1.5 M$_{\odot}$ and radius
1.64 R$_{\odot}$ we find a break-up period of 0.19~d, yet the models by \cite{landin2023} yield a convective turnover time below 0.01~d, implying that these stars can never reach the break-point in Rossby number beyond which saturation takes place.   
In other words, these stars can never become saturated, not to mention supersaturated.

\begin{figure}  [bht]
\centering
 \includegraphics[width=8cm]{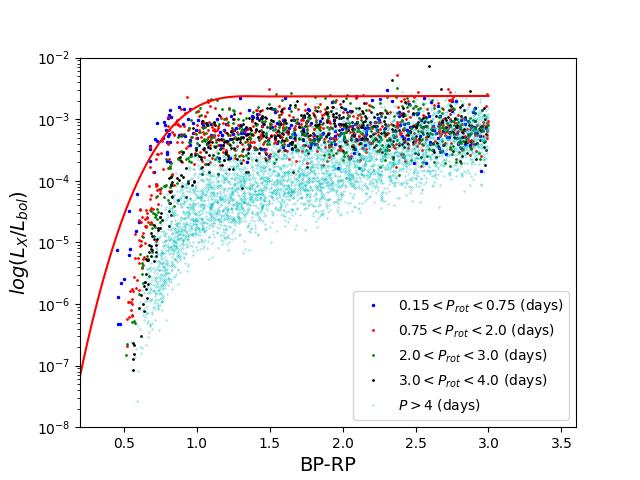}
  \vspace{-0.cm} 
   \caption{   \label{simulpop} Toy model simulation of the L$_X$/L$_{bol}$ ratio vs. BP-RP color for the eROSITA detected  cool star population. See text for details}
  \end{figure}  

\subsection{The cool star eROSITA population at large}
\label{sim_pop}
With the results obtained above, it is now possible to understand the color and X-ray activity distribution of the cool star
population in an X-ray survey such as eROSITA.  While a detailed population simulation and study would go far beyond the scope of the present
paper, we wish to demonstrate this possibility with a ''toy model'', that encompasses all the relevant features but is not meant to
model the specific eROSITA population of cool star X-ray sources.   For this purpose we perform a Monte Carlo simulation of 
a population of  MS cool stars with
uniformly distributed distances from the Sun in a spherical volume out to 100~pc and BP-RP colors again uniformly distributed; furthermore 
we assume that the statistical distribution
of rotation periods is known which we again assume to be uniform.
Since the BP-RP color determines (more or less) stellar mass, radius and bolometric luminosity,  we can then compute
the break-up period from Eq.~\ref{def_per_break} , the convective turnover time from Eq.~\ref{def_tau_conv}, and the L$_X$/L$_{bol}$ ratio 
from  Eq.~\ref{def_act_rossby_1}  or Eq.~\ref{def_act_rossby_2} for the whole population.
Since L$_{bol}$ and the distance of each source are known, the X-ray luminosity, L$_X$, and the apparent X-ray flux f$_X$ for every source 
can be computed.  For simplicity, we assumed a uniform survey
flux limit of 1 $\times$ 10$^{-14}$ erg/cm$^2$/s for the whole survey; hence, we were able to assess whether the considered source 
is detected in the survey or not.    We performed this simple Monte Carlo simulation with 10000 trials and reject any source  with a period
below the break-up period.
In this fashion, we were able to construct a simulated  L$_X$/L$_{bol}$ versus Rossby number diagram displayed in Fig.~\ref{simulpop} . In  Fig.~\ref{simulpop}, 
we distinguish between stars with different rotation periods using the same period bins as in  Fig.~\ref{lxlbolvscolor}. Furthermore,  the red solid curve
shown in Fig.~\ref{simulpop}  is identical to the upper envelope to the observed distribution shown in Fig.~\ref{lxlbolvscolor}, indicating that it has not been
derived from the simulated data.   Thus, as is clear from Fig.~\ref{simulpop}, our rather simplistic
toy model already reproduces the essential features of the eROSITA observed cool star population, suggesting that the relevant physics has  been
incorporated in the model.

\section{Conclusions}

Using TESS short-cadence photometry of eROSITA detected cool stars, we determined periods for more than
14000 objects with consistently measured X-ray fluxes. As a result, we have created the largest available sample to date of cool stars
with measured X-ray luminosities and rotation periods. Several comparisons with independently determined periods 
show that the percentage of incorrectly determined periods must be low, despite the rather massive contamination 
of TESS data by instrumental effects in some instances.   We show that a key element for the interpretation of the X-ray data is the convective 
turnover time required to compute Rossby numbers.    Our determinations of convective turnover times extend over
the full range of late-type stars and yield results that are very much consistent with theoretical model calculations; however, they 
extend only to early F-type stars.   To compute Rossby numbers,  we used the classical broken
power-law description between fractional X-ray luminosity, L$_X$/L$_{bol}$, and the Rossby number, but we
also introduced a smooth parametrization of the period-activity relations observed at X-ray wavelengths. 
Our new data confirm previously determined relations between activity as measured though 
the L$_X$/L$_{bol}$ ratio and Rossby number, while for the ''earlier'' late-type stars (specifically of types F and early G),
the stellar sample presented in this paper enables a detailed investigation of the period-activity relationships in these 
stars for the first time at X-ray wavelengths.   Both our empirical modeling as  well as theory suggest very low turnover times, mainly due 
to the small sizes of convection zones in these stars. This result implies that
extremely rapid rotators can never reach the saturation level as observed in stars of type K and M and, therefore, they 
remain ''low-activity'' stars despite their extremely rapid rotation.     Finally, we demonstrate that a
rather simple toy model can explain the observed activity distribution (measured through
L$_X$/L$_{bol}$) with respect to color (measured through {\it Gaia} BP-RP color) 
for the eROSITA-detected cool star X-ray population. We argue that the observed drop of activity by three orders
of magnitude over a narrow color range between 0.6 $<$ BP-RP $<$ 1.0 results from the properties of the
convective turnover times in these stars.

\begin{acknowledgements}
 
 We thank our referee for the careful scrutiny of our paper and the many suggestions which
 let us considerably improve our paper.
This work is based on data from eROSITA, the primary instrument aboard SRG, a joint
Russian-German science mission supported by the Russian Space Agency
(Roskosmos), in the interests of the Russian Academy of Sciences 
represented by its Space Research Institute (IKI), 
and the Deutsches Zentrum f\"ur Luft- und Raumfahrt (DLR). 
The SRG spacecraft was built by Lavochkin Association (NPOL) 
and its subcontractors, and is operated by NPOL with support from
IKI and the Max Planck Institute for Extraterrestrial Physics (MPE). 
The development and construction of the eROSITA X-ray instrument was
led by MPE, with contributions from the Dr.\ Karl Remeis Observatory 
Bamberg \& ECAP (FAU Erlangen-N\"urnberg), the University of Hamburg 
Observatory, the Leibniz Institute for Astrophysics Potsdam (AIP), 
and the Institute for Astronomy and Astrophysics of the University 
of T\"ubingen, with the support of DLR and the Max Planck Society. 
The Argelander Institute for Astronomy of the University of Bonn and 
the Ludwig Maximilians Universit\"at Munich also participated in the science 
preparation for eROSITA.
The eROSITA data used for this paper were
processed using the eSASS/NRTA software system developed by the
German eROSITA consortium. 
This work has made use of data from the European Space Agency (ESA)
mission {\it Gaia} (\url{https://www.cosmos.esa.int/gaia}), processed by
the {\it Gaia} Data Processing and Analysis Consortium (DPAC,
\url{https://www.cosmos.esa.int/web/gaia/dpac/consortium}). Funding
for the DPAC has been provided by national institutions, in particular
the institutions participating in the {\it Gaia} Multilateral Agreement.
This research has also made use of the SIMBAD database,
operated at CDS, Strasbourg, France.
This paper includes data collected with the TESS mission, obtained from the MAST data archive at the Space 
Telescope Science Institute (STScI). Funding for the TESS mission is provided by the NASA Explorer Program. STScI is 
operated by the Association of Universities for Research in Astronomy, Inc., under NASA contract NAS 5?26555.

\end{acknowledgements}

\bibliographystyle{aa}
\bibliography{bibfile}

\clearpage
\newpage

\begin{appendix}

\section{Three specific light curve examples:  AV~Dor, AB~Dor,  BD-11 1208}
\label{app:sec_abavdor}

\begin{figure}  [h]
\centering
 \includegraphics[width=8cm]{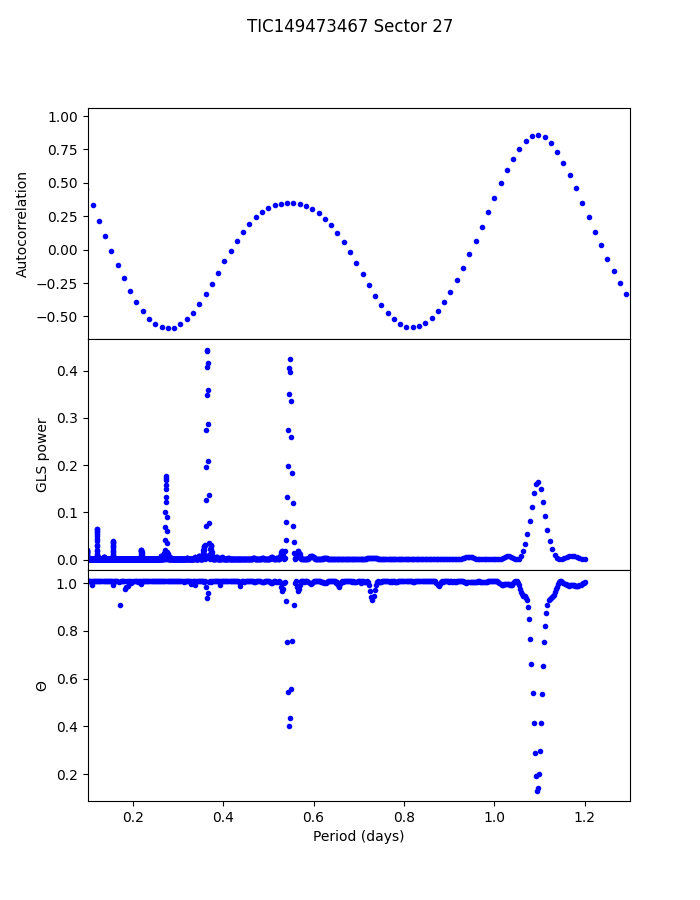}
   \caption{Period analysis for TESS data in sector 27 for AV~Dor:
   Top panel: Autocorrelation vs. period. Medium panel: GLS power vs. period. Bottom panel: $\Theta$ vs. period.
}
\label{app:avdor_glsauto}
\end{figure}

In Appendix~\ref{app:sec_abavdor} we study how our period determination scheme works on well-known stars with
known period properties to explore the limits of what can be done.
We therefore consider some specific exemplary
source types that give rise to strong periodic signals, namely, the eclipsing binary AV~Dor,
the ultra-active star AB~Dor with strong spot-induced modulations, and the multi-period star BD-11 1208.    
The nature of these
sources is well known and these examples were chosen to demonstrate 
the difficulties encountered in period determinations in general and especially in active 
stars; note that eclipsing binaries are usually also active when one or or both system components
are of late spectral type.

\subsection{AV~Dor: An eclipsing binary} \hfill \\[.5em]
\label{sec_avdor}

\begin{figure}  [ht]
\centering
 \includegraphics[width=8cm]{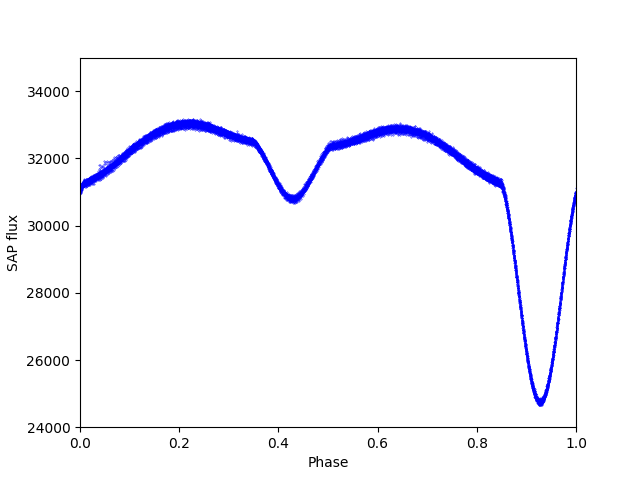}
   \caption{TESS light curve from sector 27 for AV~Dor folded with the binary period of 1.09~days; see text for details.}
\label{app:avdor_phase}
\end{figure}

AV~Dor (= TIC~149473467, spectral type F0) is a close binary with a period of 1.09~days, the TESS data of which have been analyzed by
\cite{justesen2021}.   Because of its southern location AV~Dor is frequently observed by TESS, and here we consider
the TESS data obtained in sector 27. In Fig.~\ref{app:avdor_glsauto} we show the results of our period analysis of these
data:  in the top panel we show the measured autocorrelation function, in the medium panel the GLS periodogram
and in the bottom panel the $\Theta$-statistics in the phase dispersion minimization (for details see \citealt{stelling1978}) in the period range
0.1~-~1.3~days.   As apparent from Fig.~\ref{app:avdor_glsauto}, the autocorrelation function shows two peaks at 
0.545~days and at 1.09~days, the $\Theta$ statistics has two minima at the same periods.  The GLS power spectrum shows
several maxima, yet the maximum with the largest power has a negative autocorrelation, and the period with
the largest autocorrelation and smallest $\Theta$ value has only the fourth largest power.   The period 1.09~days is
the correct period as evidenced by the folded TESS light curve of AV~Dor displayed in Fig.~\ref{app:avdor_phase}, which
shows two eclipses (at phases of about 0.45 and 0.9) as well a photometric ''wave'', due to activity phenomena on both stars 
and the maxima of which are separated by a phase difference of 0.4. 

\subsection{AB~Dor: An ultrafast rotator} \hfill \\[.5em]
\label{app:sec_abdor}

We now consider the prototypical nearby ultra-active star AB~Dor, which is also extensively observed by TESS
due to its favorable southern location near the southern ecliptic pole;
eROSITA and TESS data of AB~Dor have already been presented by \cite{schmitt2021}.
The rotation period of AB Dor is well known, with $P_{rot}$ = 0.51428~days \citep{ioannidis2020} it
executes more than 50~rotations in a typical TESS sector and may exhibit photometric 
modulations of more than 10\%, leading to an
extremely ''sharp'' signal in Fourier space; a very detailed study of TESS data of
AB~Dor has been presented by \cite{ioannidis2020}.   

\begin{figure}  [h]
\centering
 \includegraphics[width=8cm]{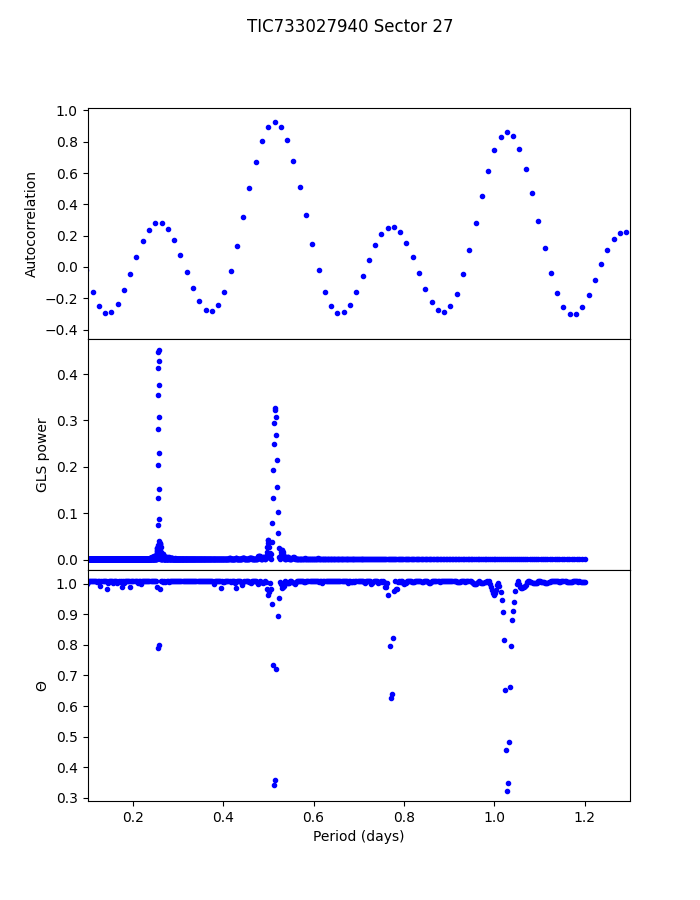}
   \caption{Period analysis for ultrafast rotator AB~Dor with TESS sector 27 data.
   Top panel:  Autocorrelation function vs. period. Medium panel: GLS power vs. period. Bottom panel: $\Theta$ vs. period.} 

\label{app:abdor_glsauto}
\end{figure}

In Fig.~\ref{app:abdor_glsauto} we plot (analogously to Fig.~\ref{app:avdor_glsauto}) the GLS periodogram (medium panel) for the TESS data on AB~Dor in sector 27; the top panel shows the
 autocorrelation of the data, the bottom panel the $\Theta$-statistics for the phase dispersion minimization. 
 The GLS periodogram in Fig.~\ref{app:abdor_glsauto} does show the expected
sharp Fourier peaks: one at the known rotation period of $P_{rot} = 0.51428$ days and the other peak appears
at half this period as an alias, yet with even more Fourier power than the correct period peak.   In the
autocorrelation of the phased data (shown in the top panel of Fig.~\ref{app:abdor_glsauto}), we can  also recognize two
much broader peaks at the same periods as in the periodogram, however, the phased light curve autocorrelation
at the correct period is larger than that of the aliased period.  Similar consideration apply to the $\Theta$-statistics
for the phase dispersion minimization; the $\Theta$-statistics is smallest for the correct period, but there
are aliases at half the correct period as well as integer multiple of these periods.

\begin{figure}  [h]
\centering
 \includegraphics[width=8cm]{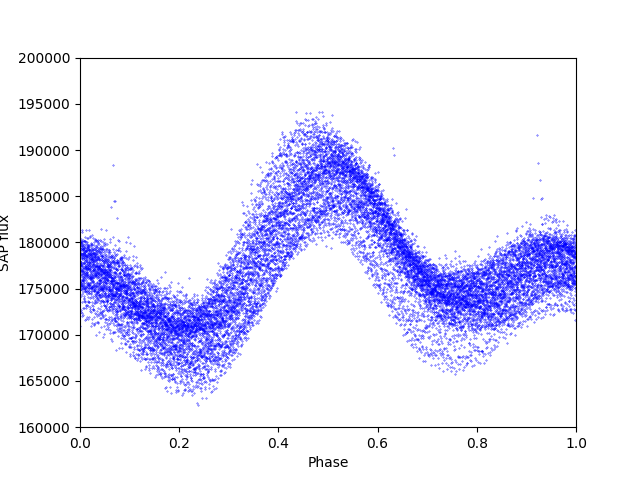}
   \caption{TESS sector 27 light curve for AB~Dor folded with the binary period of 0.51428~days.}
\label{app:abdor_phase}
\end{figure}

It is instructive to examine the TESS data folded with the rotation period as plotted in Fig.~\ref{app:abdor_phase}.
We would  immediately notice the substantial photometric variability of AB~Dor even outside obvious flares,
and the relative peak-to-peak amplitude is almost 14\%.   Furthermore, we can make note of the two flux minima, namely, phases when the surface spot concentration is largest:  one, at about $\phi \approx $ 0.25,
and another one at $\phi \approx $ 0.8 (i.e., almost exactly on the other side of the star).

\subsection{BD-11 1208: A multi-period star} \hfill \\[.5em]
\label{sec_bd11}

\begin{figure}  [h]
\centering
 \includegraphics[width=8cm]{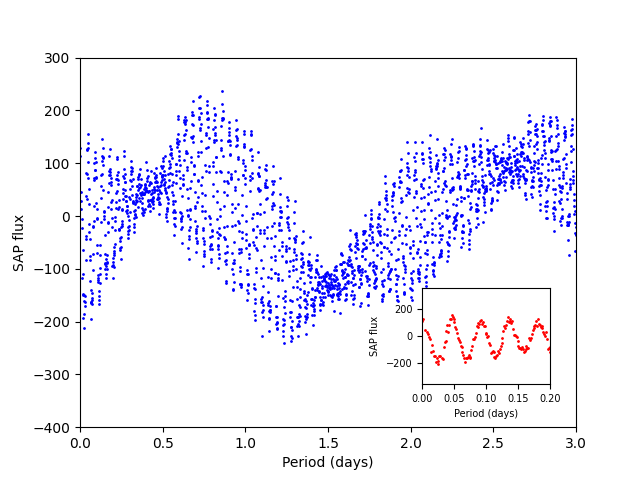}
\caption{TESS  sector 32 light curve for BD-11 1208. See text for details.}
\label{app:bd111208}
\end{figure}

\begin{figure}  [h]
\centering
 \includegraphics[width=8cm]{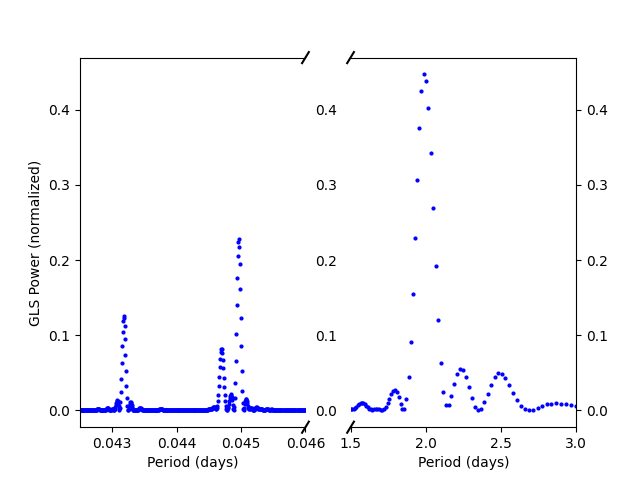}
\caption{GLS periodogram of TESS sector 32  light curve for BD-11 1208. See text for details.}
\label{app:bd111208gls}
\end{figure}

\noindent We finally considered a star with multiple periods, namely, the star BD-11~1208 (Gaia DR3 3009996545138727936 or
TIC~201372298), which appears to have received rather little attention so far; according to SIMBAD it is of spectral type A; this is consistent with the {\it Gaia} color information.   In Fig.~\ref{app:bd111208} we show the first few days of the
TESS data obtained in sector 32 of this star, with the inset zooming in on the first 5 hours of this data set,  and in Fig.~\ref{app:bd111208gls}, 
we show the GLS periodogram for all of the TESS data for BD-11~1208.  As shown in Fig.~\ref{app:bd111208gls}, there are
two rather significant periods apparent in the TESS time series, a shorter period of $\sim$ 1 hour is clearly visible in the modulations in
the inset of Fig.~\ref{app:bd111208}, along with a longer period of $\sim 2 $ days.  

It is obvious that the shorter period cannot possibly
be a rotation period, it rather appears to be a pulsational period of the A-type star.   The longer period could be interpreted as the
stellar rotation period, alternatively, it might be the rotational or orbital period, if we interpret BD-11 1208 as a binary system,
an interpretation suggested by the large RUWE value derived by  {\it Gaia}.   Finally we note that
\cite{gaia2023} list this source in their non-single star catalog 
and quote a period of (2.5427157 $\pm$ 0.000097)~d 
with an almost circular orbit for this object.   As is clear from  Fig.~\ref{app:bd111208gls} , the main peak of the GLS periodogram 
is at 1.987~d, while at the period reported by \cite{gaia2023} merely a minor aliasing peak is located.  We further note
that the correlation and phase dispersion analyses strongly favor the two day period and conclude that a period of 2.5~d is
inconsistent with the TESS photometry.   At any rate, BD-11 1208 shows the pitfalls of photometric period analysis.

\section{Period determination procedure}
\label{app:sec_per_det}

The examples of AV~Dor, AB~Dor and BD-11~1208 presented in Appendix~\ref{app:sec_abavdor}
demonstrate the difficulties encountered in period determinations 
in general and especially in active 
stars.   Clearly, the GLS periods (for short-period systems) are very well determined (cf., Figs.~\ref{app:avdor_glsauto} 
and~\ref{app:abdor_glsauto}) and are therefore 
to be preferred; on the other hand, as shown by the
cases of AV~Dor and AB~Dor, the period associated with the largest GLS power need not necessarily 
be the correct period, and period aliasing is a well-known and severe problem especially for Fourier-type methods, 
but also for other period search algorithms.  In the cases of AV~Dor and AB~Dor, 
the correct periods are well known and we can properly interpret an apparently ''anomalous''  light curve 
as in Fig.~\ref{app:abdor_phase}; yet
if we only had data from TESS sector 27 at our disposal, we would encounter interpretational 
difficulties.  It appears that for active stars the GLS periods appear to be
quite trustworthy, yet it is always important to check possible aliasing.  We therefore compute for all our periods the
values of the autocorrelation correlation function and the $\Theta$ statistics to safeguard against 
aliasing; we do never accept
GLS periods with negative values for the autocorrelation function 
(as is the case for the maximal GLS power in the case of AV~Dor), yet we can
never be absolutely sure not to be fooled by aliasing effects. 

Obviously,  short-period systems execute many rotations in a given TESS sector and
therefore produce clear and sharp signals in Fourier space as shown in  Figs.~\ref{app:avdor_glsauto} and~\ref{app:abdor_glsauto}, with the
Fourier peaks being much narrower than the peaks found in the autocorrelation function.
On the other hand, a low activity star like our Sun with a rotation period of about 25~days would not be detected
with our procedures because of the limited sampling period and typically low variability amplitude. 
Even a moderately active star with a rotation period of, say, 10~days would execute only three rotations in a given TESS sector, which
would then naturally lead to much broader signals in the resulting periodograms; also such a period would
already come rather close to the instrumental period of 12-13 days.

For these reasons we decided to carry out independent period searches for short and
long periods with slightly different procedures, since it is clear that short-period systems need much 
finer period sampling to accurately determine
the peak in a GLS periodogram.   In this context, short periods are assumed to be between 0.02~days and 0.6~days.  
Obviously, a period of 0.02~days is well below the period we would expects for a stellar rotation period 
and close to or below the break up period, so any period that low cannot be 
a rotation period.   Long periods are then all periods longer than 0.6~days.

As discussed above, for us the GLS period is our method of choice and we search all our sample stars on our period
grid for periods with GLS peak power above 0.1; our experience is that true periods typically do reach
that value.  However, we found it both advantageous and necessary to also consider other period indicators, and 
in particular, we always demand that the autocorrelation at the peak value exceed 0.1 to be acceptable
as a valid period measurement and we always also examine the phased light curves using the phase dispersion minimization
approach  developed by \cite{stelling1978}.

\subsection{Determination of long periods} \hfill \\[.5em]
\label{app:sec_longper}

In our period determination scheme, long periods are periods in the range of 0.6~ - $\approx$ 16~days.
As demonstrated in Appendix~\ref{app:sec_abdor}, we use multiple period determination schemes to reduce 
the  very frequently encountered difficulties by aliasing effects.   For the longer periods,
fewer and fewer cycles are covered and periods of about 12~days and more
are heavily affected by a variety of instrumental effects.
We use periodogram, autocorrelation and phase dispersion
minimization analysis and illustrate our period search 
procedures for two TESS light curves, a ''good'' light curve
and a ''poor'' light curve, respectively.   

\begin{figure}  [h]
\centering
 \includegraphics[width=8cm]{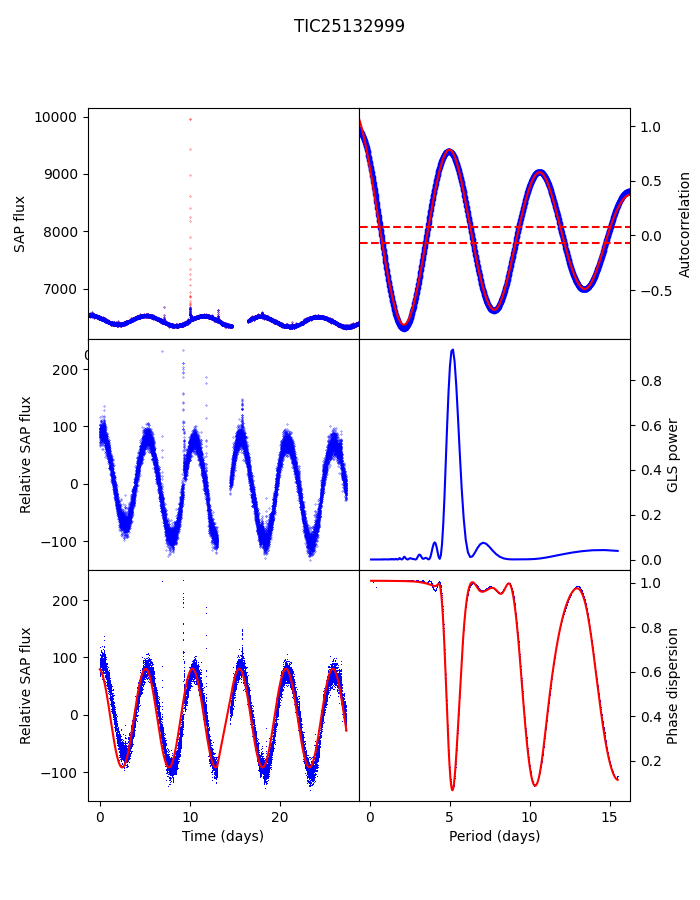}
   \caption{TESS light curve analysis for the TESS data (in sector 2) for TIC~25132999. The data are shown in
   the left panels, the period results in the right panels; see text for details.}
   \label{app:samp_anal_1}
  \end{figure}

A ''good'' light curve and its analysis is shown in Fig.~\ref{app:samp_anal_1} for the
case of TIC~25132999, which refers to the star EXO 040830-7134.7, a dMe star extensively studied already
with EXOSAT \citep{vandenwoerd1989}.  In the left panel of Fig.~\ref{app:samp_anal_1} we plot (top panel) the TESS
SAP flux, which shows a modulated light curve, on top of which there are a number of flares; the data
gap after about 13~days is also visible.  In middle left panel of Fig.~\ref{app:samp_anal_1} we show 
the rectified light curve, used for all period analyses; note that no attempt was undertaken to remove the flares
from the recorded light curve.  In the right panels of Fig.~\ref{app:samp_anal_1} we show the analysis
results, i.e., the autocorrelation function vs. period (top right panel), the GLS power (medium right panel)
and the $\Theta$-statistics of the phase dispersion minimization (bottom right panel).  While autocorrelation and
phase dispersion show a multitude of peaks, the GLS periodogram shows a single strong peak at a period
of 5.18~d, which we interpret as rotation period of TIC~25132999; the GLS best period sine wave corresponding to
this period yields a surprisingly good reproduction of the observed light curve, as evidenced in the bottom-left panel of
Fig.~\ref{app:samp_anal_1}.

As demonstrated in Fig.~\ref{app:samp_anal_1}, in the specific case of TIC~25132999 both the autocorrelation and 
phase dispersion curves
show a periodic structure, whereas the GLS periodogram yields a single large peak.
While it is possible to create circumstances where all of the GLS power is
shifted to half the true period (for example, by locating two identical spots
180~degrees apart on the stellar surface as in the case of AB~Dor), we argue that in the specific case of
the TESS data for TIC~25132999 in TESS sector 2 the GLS period is the correct period which we choose for
our further analysis;
incidentally, TIC~25132999 is also contained in sector 1 and our period analysis yields
identical results. 

\begin{figure}  [h]
\centering
 \includegraphics[width=8cm]{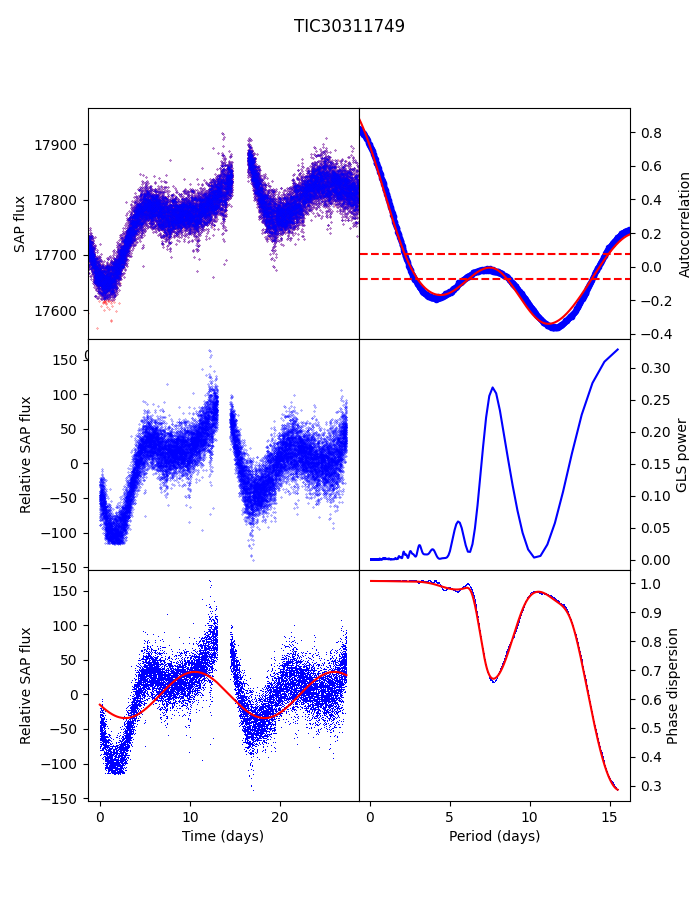}
   \caption{TESS light curve analysis for the TESS data (in sector 2) for TIC~30311749. The data are shown in
   the left panels, the period results in the right panels; see text for details.}
   \label{app:samp_anal_2}
  \end{figure}

We next show an example of a ''poor'' light curve for the case of TESS data for TIC~30311749 obtained 
in sector 2; incidentally, TIC~30311749 is identical to HD 268670 (K0e, a spectroscopic binary), namely,
an active star. In  Fig.~\ref{app:samp_anal_2}, we show our analysis of the TESS data for TIC~30311749 
in an identical fashion as the TESS data for TIC~25132999 shown in  Fig.~\ref{app:samp_anal_1}.
The TESS light curve of TIC~25132999 shows the usual gap after about 13~days; furthermore, there are
modulations, which are (presumably) in part intrinsic and in part instrumental.   The mean
levels before and after the 13~days break are different, and the rapid variations near the break
are quite odd.   The period analysis yields a seemingly significant period near 15~days, which,
however, was introduced by our light curve normalization procedures.  In addition, the GLS periodogram
shows a peak near 8~days with a GLS power of more than 0.25.   While we would normally accept periods
with such power values and while this value
would actually be a reasonable rotation period for an active K dwarf, we cannot recognize a clear autocorrelation signal; 
therefore, we conclude that in this specific case of TIC~30311749 
no trustworthy period assignment can be made on the basis of the available data.

\begin{figure}  [h]
\centering
 \includegraphics[width=8cm]{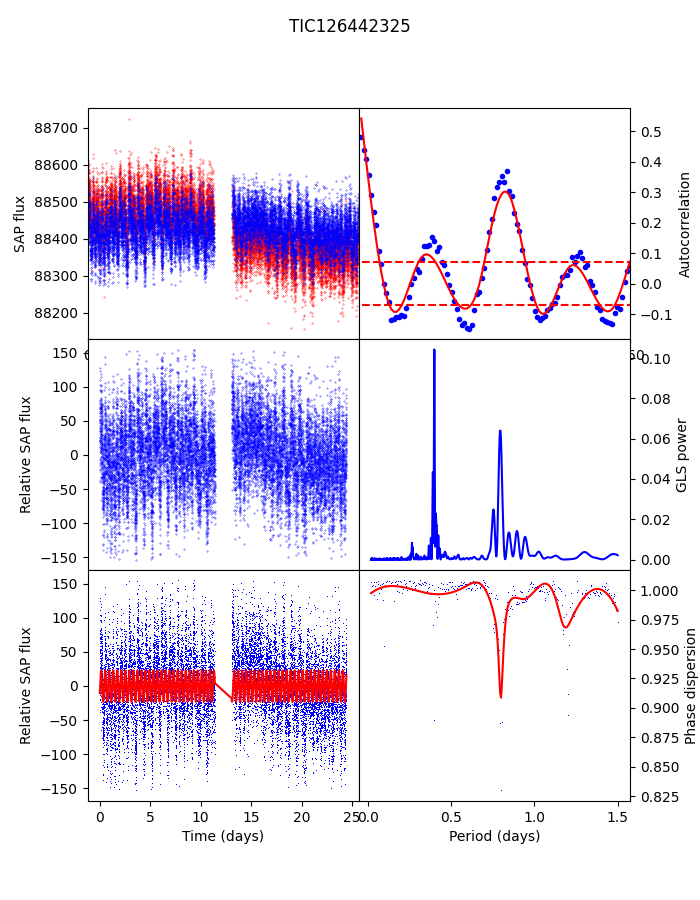}
   \caption{TESS light curve analysis for the TESS data (in sector 7) for TIC~126442325. The data are shown in
   the left panels, the period results in the right panels. See text for details.}
   \label{app:samp_anal_3}
  \end{figure}

\subsection{Determination of short periods} \hfill \\[.5em]

\label{app:sec_shortper}

We now turn to the period determination for short periods. Two examples for short period systems with very good light curves
have already been produced for the cases of
AV~Dor (see Fig.\ref{app:avdor_glsauto}) and AB~Dor (see Fig.~\ref{app:abdor_glsauto}).   In the following, we consider
a marginal case: that of TIC~126442325, which  is identical with the F0V star HD~62707.   The results of our
period analysis of the TESS data in sector 7 of this source are shown in Fig.~\ref{app:samp_anal_3} in a fashion completely
analogous to  Figs.~\ref{app:samp_anal_1} and~\ref{app:samp_anal_2}.  We can recognize  from  Fig.~\ref{app:samp_anal_3}
a periodicity albeit at rather low levels with a peak-peak amplitude at around the 0.2\% level.   The main
GLS peak at 0.4~days is barely above our cutoff power level at 0.1, the same applies to the autocorrelation value;
at twice the period, we can see greater autocorrelation and phase dispersion signals, while the GLS power
drops below 0.1.   

In the {\it Gaia} catalog releases, we find a stellar radius of 1.35 $R_{\sun}$ (in DR2) and
a spectral broadening velocity v$_{broad}$ =  81.4 $\pm$ 2.6 km/s (in DR3) for HD~62707; unfortunately the actual spectra
from Radial Velocity Spectrometer (RVS) on board {\it Gaia}
are not available, so no independent analysis can be carried out.
If we now interpret v$_{broad}$ as a v sin(i) measurement with $i$ = 90$^{\circ}$, we find a period of 0.84~days, which would 
actually reasonably fit to the aliased main GLS peak at 0.4~days.  Thus we may actually see the rotational signal of
HD~62707 in the TESS data, however, TIC~126442325, does clearly represent a limiting case.
The phase dispersion statistics $\Theta$  does not show a very clear signal, therefore we refrain from assigning a period to 
TIC126442325, although the TESS autocorrelation data do suggest some periodicity.

\subsection{Final period assignments}
\label{app:sec_fin_per}

In the previous sections we have tried to elucidate the quality range of TESS light curves for the counterparts of eROSITA X-ray sources.
Obviously, in the end we have to arrive at a single (hopefully rotation) period for any given star.  For each X-ray source, we might have 
access to period measurements in a given sector (which might possibly be in disagreement) using the three period assessment methods, whereas for
a number of sources, we have period measurements in different sectors (for some sources we have coverage in more than 20 sectors); finally, we might encounter successful period measurements both in the long and short period ranges.
To deal with this situation, we developed an empirical period classification
scheme that assigns a final period and a period quality flag to every star.

As discussed earlier in this work, our period determination scheme is primarily based on the GLS periodogram power (as defined by \citealt{zech2009}); any 
successful measurement must yield a GLS power of at least 0.1 and such a measurement is assigned a grade of 1; any maximal GLS power below this value is not considered  as successful and is assigned grade 0 and ignored in the following analysis.
If a GLS period (with a power above 0.1) is ''confirmed'' by a measurement of phase dispersion minimization (PDM) and/or autocorrelation (AC) 
the grade is assigned 2 or 3, if successful period measurements (consistent to within 10\%) are available
for GLS+PDM and GLS+AC combinations respectively. The best grade assigned to a period derived from data in a single sector, grade 4, requires successful and 
consistent period determinations (to within 10\%) with all the GLS, AC and PDM 
methods. As described above, this type of analysis
was carried out separately for short periods (below 1.6~days) and long periods.

For many of our eRASS1 X-ray sources,
we have TESS coverage in more than one sector, however, there are also many sources that were
observed by TESS only in a single sector.   Clearly, repeated and consistent period determinations in
different sectors increase our confidence in the correctness of the period determined for a given star.
We therefore increase the grade assignment of consistent period measurements that are available for
a given star in several sectors.   For sources with successful period measurements in more than one sector  available, we first identified the sector with the best period quality and increased the assigned quality grade by 1;
if at least three consistent period measurements in different sectors are available; by 2 if between 4 and 8
consistent period measurements in different sectors are available; and by 3, if more than 8  
consistent period measurements in different sectors are available. Thus, the maximum possible grade is
7, implying that there are at least nine consistent high quality period measurements available for the given 
source.

Finally, we needed to merge the short and long period measurements.    In the majority of cases (16004), we obtained
only a successful long period measurement, in the minority of cases (1106) we obtained
only a successful short period measurement, but there are some cases (913) with both long and short period measurements.
Almost half of the sources in both categories have periods in the overlap region; namely, the derived periods are consistent with
each other.  In 116 (out of 913) cases, the derived long period is an alias of the short period and in less than 1\% of the cases, the long period is less than 90\% of the short period, a clear signature of insufficient period sampling in the GLS calculation.  Finally we have 388 remaining cases, where
the ratio between the long and short period measurements is often very large; namely, in most cases, due to the fact that
a successful short period was determined, while a long period was found near 13.6~days or near 6.25~days, which we interpret
as an alias introduced by the TESS data sampling.   In these overlapping cases, we therefore adopted the derived short period, except in
cases where the long period is an alias of the short period (where we considered the longer period as the more likely one).

In the selected TESS sectors, we analyzed more than 69000 individual TESS light curves, which belong to 22461 different individual sources.   
For 14004 sources we managed to obtain valid period measurements with grades of 3 and higher;
the distribution of the grade quality is shown in Table~\ref{tab2}.  Only 11731 of the derived periods have quality assignments greater than 3; that does not 
imply that the periods for the remaining 2273 sources are wrong, however, the rate of incorrect period determinations
among these sources is expected to be  higher. An exemplary listing of our results is given in Table~\ref{tab1}, the full table is available in the online material.

\begin{center}
\begin{table}
\centering
\caption{\label{tab2} Quality distribution for successful period determinations.} 
\begin{tabular}{  l r}\\
\hline
\hline
 Period grade& Number\\
\hline
 3 & 2273\\
 4 & 7542\\
 5 & 3063\\
 6 & 670\\
 7 & 456\\
\hline
\end{tabular}                   
\end{table}
\end{center}

\section{A catalog of TESS periods for eROSITA X-ray sources}

In the fashion described in Appendix~\ref{app:sec_per_det}, we analyzed a total of
65157 individual TESS sector light curves which refer to 18023 unique sources.
An exemplary listing of our results is given in Table~\ref{tab1}, the full table is available in the online material 
where we provide a list with all our period determinations, containing TIC number, derived period (in days), the estimated
period error (also in days) as well as the grade assigned by us to the respective period determination as well as two flags, the 
full table is available only electronically.

\end{appendix}

\end{document}